\documentstyle[floats,prl,aps,epsf,eqsecnum,preprint]{revtex}
\begin{document}

\title{Absorption and Emission Spectra of an higher-dimensional 
Reissner-Nordstr\"{o}m black hole} 
%\preprint{MCTP-01-19}
\author{Eylee Jung\footnote{Email:eylee@kyungnam.ac.kr} and
D. K. Park\footnote{Email:dkpark@hep.kyungnam.ac.kr 
}, }
\address{Department of Physics, Kyungnam University, Masan, 631-701, Korea}

\maketitle

%\date{\today}
\maketitle
\begin{abstract}
The absorption and emission problems of the brane-localized and bulk scalars
are examined when the spacetime is a $(4+n)$-dimensional Reissner-Nordstr\"{o}m
black hole. Making use of an appropriate analytic continuation, we compute the 
absorption and emission spectra in the full range of particle's energy. For the case of
the brane-localized scalar the presence of the nonzero inner horizon 
parameter $r_-$ generally enhances the absorptivity and suppresses the emission 
rate compared to the case of the Schwarzschild phase. The low-energy absorption cross
section exactly equals to $4\pi r_+^2$, two-dimensional horizon area. The effect of 
the extra dimensions generally suppresses the absorptivity and enhances the emission
rate, which results in the disappearance of the oscillatory pattern in the total 
absorption cross section when $n$ is large. For the case of the bulk scalar the effect
of $r_-$ on the spectra is similar to that in the case of the brane-localized scalar.
The low-energy absorption cross section equals to the area of the horizon 
hypersurface. In the presence of the extra dimensions the total absorption 
cross section tends to be inclined with a positive slope.  
It turns out that the ratio of the 
{\it missing} energy over the {\it visible} one decreases with increase of $r_-$.
\end{abstract}

%---------------------------------------------------------------------------
\newpage
\section{Introduction}
Black hole is an important physical system in a sense that we can 
formulate, test, and explore the quantum gravity, the untimate theory of 
physics, within this system. Although there is some doubt in the 
physical realization in the context of the information loss 
problem\cite{hawk76,pre92}, the most well-known quantum process in the 
black hole physics is a definitely Hawking radiation\cite{hawk74,hawk75}. 
Recently, therefore, there are various attempts to verify the Hawking effect
by exploiting the analogue systems\cite{unruh04}. 

According to the Hawking formula the power spectrum, {\it i.e.} the energy 
emitted per unit time, is given by
\begin{equation}
\label{hawking1}
\Gamma_D = \omega \frac{\sigma_{abs}(\omega)}{e^{\frac{\omega}{T_{H}}} \pm 1}
\frac{d^{D-1} k}{(2 \pi)^{D-1}}
\end{equation}
where $D$ is a spacetime dimensions and $T_H$ is an Hawking temperature. The
denominator is a Planck factor, and the upper and lower signs correspond
to the fermion and boson, respectively. The appearance of the Planck factor 
in the emission spectrum indicates that the black hole is a thermal object.
The numerator $\sigma_{abs}(\omega)$ is a total absorption cross section
which is a sum of the partial absorption cross section $\sigma_{\ell}(\omega)$;
\begin{equation}
\label{totalabs}
\sigma_{abs}(\omega) = \sum_{\ell} \sigma_{\ell}(\omega).
\end{equation}
Although the summation index $\ell$ represents, in general, the set of all
quantum numbers, this becomes the orbital quantum number for the spherically
symmetric black holes, which is the case in this paper. The partial absorption
cross section $\sigma_{\ell}(\omega)$ in $D$-dimensions is expressed in the 
form\cite{gub97-1}
\begin{equation}
\label{partialabs}
\sigma_{\ell}(\omega) = \tilde{A}_H
\frac{2^{D-4} \Gamma^2 \left(\frac{D-1}{2}\right)}{\pi (\omega v r_H)^{D-2}}
\frac{(2 \ell + D - 3) (\ell + D - 4)!}{(D-3)! \ell !}
{\cal T}_{\ell} (\omega)
\end{equation}
where $\tilde{A}_H$ is an area of the horizon
\begin{equation}
\label{area1}
\tilde{A}_H = \frac{2 \pi^{\frac{D-1}{2}}}{\Gamma \left(\frac{D-1}{2}\right)}
              r_H^{D-2}.
\end{equation}
The factor $v$ in Eq.(\ref{partialabs}) is a velocity factor defined 
$v = \sqrt{1 - m^2 / \omega^2}$, where $m$ is a particle mass. 
${\cal T}_{\ell} (\omega)$ is a transmission coefficient, which is a 
transmission probability for the incident ingoing wave shooted from the asymptotic
region. Since this factor should be dependent on the potential generated 
by the horizon structure, it encodes a valuable information for the 
nature of spacetime. Since, furthermore, ${\cal T}_{\ell} (\omega) \neq 1$
in general, this factor makes the black hole distinct from the black body.
In this reason the factor is often called `greybody' factor.

As Eq.(\ref{hawking1}) indicates, the absorption cross section plays an 
important role for understanding the emission spectrum. Many computational 
techniques for the calculation of the absorption cross section were 
developed about three decades 
ago\cite{teuk72,sta73,cart74,ford75,page76,unruh76,sanc78}. The computational 
procedure most peoples adopted is a matching between the near-horizon and
the asymptotic solutions. Following this procedure, Unruh\cite{unruh76} derived
the low-energy absorption cross section analytically for the massive scalar and
Dirac fermion in the $4d$ Schwarzschild background. For the scalar case the 
low-energy cross section $\sigma_S$ coincides with $\sigma_S = A_H / v$ when 
we take a s-wave, where $A_H$ is an area of horizon and $v$ is a velocity 
factor. This means the low-energy cross section for the s-wave is equal to 
the horizon area in the massless limit. It was also found by Unruh that the 
low-energy cross section $\sigma_F$ for the massive Dirac fermion bacomes 
$\sigma_F = \sigma_S / 8$ when we take $j=1/2$, where $j$ is a total 
angular momentum. Recently, this ratio factor is generalized to 
$2^{(n-3) / (n+1)}$ in 
the $(4+n)$-dimensional Schwarzschild phase. 

The fact that the low-energy absorption cross section for the s-wave massless
scalar equals to the horizon area is generally proved in the higher-dimensional
asymptotically flat and spherically symmetric black holes\cite{das97}. This 
universal property was re-examined in the $p$-brane-like 
object\cite{emp98,park00,jung03} and non-spherically symmetric black 
hole\footnote{It is unclear, however, at least for us that non-spherically 
black hole such as BTZ black hole still holds the universality. For example,
the authors in Ref.\cite{birm97} choosed the constant $c$ in Eq.(36) of 
their paper to hold the universality. But this constant $c$ is not a free 
parameter. Thus, we think this should be uniquely fixed using the basic 
principle of physics. This will be discussed elsewhere.} 
such as the three-dimensional BTZ black hole\cite{birm97}. 

The absorption and emission problems for the full range of particle's energy
were also computed by applying the quantum mechanical scattering
theories\cite{sanc78,sanc98,jung04-2}. Adopting an appropriate numerical 
techniques, it was found that the total absorption cross section with respect
to the particle's energy generally exhibits a wiggly pattern, which implies 
that each partial absorption cross section has a peak in different energy 
scale. However, the total emission rate does not have the oscillatory 
pattern, which indicates that the Planck factor in general suppresses the 
contribution of the higher partial waves except s-wave. It turned out also
that the scalar mass reduces the emission rate. 

Besides the general relativity black holes are important physical laboratories
in other theoretical branches such as the string theories and the brane-world
scenarios. In string theories the attempt to understand the Bekenstein-Hawking
entropy\cite{beken73,hawk76-2}, one quarter of the horizon, microscopically 
was initiated in Ref.\cite{strom96}. Ref.\cite{strom96} used the 
five-dimensional extremal black hole by counting the degeneracy of the 
Bogomol'ni-Prasad-Sommerfield(BPS)-saturated D-brane bound states. Later, this
is extended to the certain near-extremal states\cite{horo96}. Subsequently,
the correspondence between the black hole and D-brane was focused to the
decay-rate {\it via} the Hawking radiation\cite{das96}. In Ref.\cite{mal96-1} 
five-dimensional Reissner-Nordstr\"{o}m(RN) black hole carrying three 
different electric charges was chosen to test the correspondence. 
In particular, Ref.\cite{mal96-1} showed the exact coincidence between 
the semi-classical black hole approach and D-brane approach in the 
absorption and emission rates of the neutral and charged scalars in the 
dilute gas region. These calculations were extended to the absorption and 
emission problems for the particles with arbitrary spin\cite{cve98-1}. 
However, it is not clear whether the exact agreement between these two
different approaches is maintained beyond the dilute gas region or not. 
The possibility for the disagreement was argued in Ref.\cite{hawk97} 
in the near-extremal range slightly different from the dilute gas region
although the rigorous proof was not given yet. 

The emergence for the 
TeV-scale gravity in the brane-world scenario such as the large extra 
dimensions\cite{ark98-1,anto98} or warped extra dimensions\cite{rs99-1} opens
the possibility to make tiny black holes factory in the future high-energy
colliders\cite{gidd02-1,dimo01-1,eard02-1,stoj04}. In this context it is 
important to study the absorption and emission problems of the 
higher-dimensional black holes for the future experiments. Recently, works 
along this direction were 
done\cite{jung04,kanti02-1,kanti03-1,harris03-1,jung04-3,kanti05}. Especially, 
Ref.\cite{jung04} showed that the ratio factor 
$\sigma_F / \sigma_S = 1/8$\cite{unruh76} in the four-dimensional Schwarzschild
black hole is changed into $\sigma_F / \sigma_S = 2^{(n+3)/(n+1)}$ in 
$(4+n)$ dimension. Thus this ratio factor may give an evidence for the 
existence of the extra dimensions in the future black hole experiments.

In this paper we would like to compute the absorption and emission spectra
numerically for the brane-localized scalar and bulk scalar in the 
full-range of energy when the spacetime is a $(4+n)$-dimensional 
RN black hole\cite{tang63,myers86}
\begin{equation}
\label{space1}
ds^2 = -\left[1 - \left(\frac{\tilde{r}_+}{\tilde{r}}\right)^{n+1} \right]
        \left[1 - \left(\frac{\tilde{r}_-}{\tilde{r}}\right)^{n+1} \right] dt^2
      + \frac{d \tilde{r}^2}
             {\left[1 - \left(\frac{\tilde{r}_+}{\tilde{r}}\right)^{n+1} \right]        \left[1 - \left(\frac{\tilde{r}_-}{\tilde{r}}\right)^{n+1} \right]}
      + \tilde{r}^2 d \Omega_{n+2}^2
\end{equation}
where
\begin{equation}
\label{angle-part}
d\Omega_{n+2}^2 = d\theta_{n+1}^2 + 
\sin^2 \theta_{n+1} \Bigg( d\theta_{n}^2 + \sin^2 \theta_n \bigg(
\cdots + \sin^2 \theta_2 \left( d\theta_1^2 + \sin^2 \theta_1 d\varphi^2
           \right) \cdots \bigg) \Bigg).
\end{equation}
It is well-known that the mass $M$, charge $Q$, entropy $S$ and Hawking 
temperature\footnote{Although there is a debate on the black hole radiation
temperature\cite{hooft84}, we will choose the usual definition of the 
Hawking temperature, {\it i.e.} inverse of the period, in the Euclidean spacetime 
structure.} $T_H$ of the spacetime (\ref{space1}) are
\begin{eqnarray}
\label{mqst}
M&=&\frac{n+2}{16 \pi} \Omega_{n+2} \left(r_+^{n+1} + r_-^{n+1} \right)
                                                 \\   \nonumber
Q&=&\pm \left(r_+ r_-\right)^{\frac{n+1}{2}}
    \sqrt{\frac{(n+1) (n+2)}{8 \pi}}             \\   \nonumber
S&=& \frac{1}{4} \Omega_{n+2} r_+^{n+2}
                                                 \\    \nonumber
T_H&=& \frac{n+1}{4 \pi r_+^{n+2}} \left(r_+^{n+1} - r_-^{n+1} \right)
\end{eqnarray}
where
\begin{equation}
\label{unit-sphere}
\Omega_{n+2} = \frac{2 \pi^{\frac{n+3}{2}}}
                    {\Gamma \left( \frac{n+3}{2} \right)}
\end{equation}
is the area of a unit $(n+2)$-sphere.

In this paper we will take an arbitrary $r_-$, which enables us to treat 
the Schwarzschild and extremal black holes as an unified way. Although our
paper is strongly motivated by the recent brane-world scenarios, our
treatment of the inner-horizon parameter $r_-$ may give some insight into the
black hole-D-brane correspondence beyond the dilute gas domain. 

This paper is organized as following. In section II the general properties 
of the scalar wave equation for the brane-localized scalar are examined.
The effective potential, Wronskian between two independent solutions and 
any other physical quantities are explicitly computed. 
Two solutions which are convergent in the near-horizon and asymptotic regimes
respectively are derived as an analytic form.
In section III the absorption 
and emission spectra for the brane-localized scalar 
is computed numerically by applying the quantum 
mechanical scattering theories. 
To carry out a calculation  the solution which has a convergent range 
around $r = b$ where is 
$b$ is some parameter is constructed. Using this solution the 
near-horizon solution is matched with asymptotc solution {\it via} 
the analytic continuation. 
It turns out that the presence of the nonzero inner horizon parameter $r_-$
generally enhances the absorptivity and suppresses the emission rate compared to 
the case of the Schwarzschild phase. The effect of the extra dimensions generally 
suppresses the absorption rate, which results in the disappearance of the oscillatory
pattern in the total absorption cross section. 
In section IV the general properties of the wave equation for the bulk 
scalar is examined. As in section II all physical quantities are expressed in terms
of the jost functions. The analytic near-horizon and asymptotic solutions are 
explicitly derived. 
The numerical approach for the absorption and emission spectra of the 
bulk scalar is discussed in next section. The effect of $r_-$ on the spectra is 
similar to that for the case of the brane-localized scalar. However, the universality
of the s-wave cross section makes the oscillatory pattern in the total absorption 
cross section to be maintained. A remarkable fact in the presence of the extra
dimensions is that the total absorption cross section tends to be inclined with
a positive slope. The slope seems to increase with increasing $n$.  
In section VI the ratio of the 
missing energy, {\it i.e.} emission into the bulk, over the visible one are discussed. It
turns out that this ratio factor decreases with increasing $r_-$. When $n$ is not
too large, the ratio factor is smaller than unity, which indicates that the emission
into the brane is dominant. This is consistent with the main result of 
Ref.\cite{emp00}. If, however, $n$ is very large, the emission into the bulk can 
be dominant. In section VII a brief conclusion is given.  

\section{wave equation for the brane-localized scalar}
In this section we would like to examine the various properties of the wave 
equation for the brane-localized scalar $\Phi_{BR}$ minimally coupled to the 
spacetime (\ref{space1}). Thus we should assume that $\Phi_{BR}$ is a function 
of the brane worldvolume coordinate, {\it i.e.} 
$\Phi_{BR} = \Phi_{BR} (t, r, \theta \equiv \theta_{n+1}, \varphi)$. 
Furthermore, we assume that the $3$-brane is located at
$\theta_1 = \theta_2 = \cdots = \theta_n = \pi / 2$, where 
$\theta_1, \theta_2, \cdots$ and $\theta_n$ are toroidally compactified
extra dimensions. Thus the induced metric on the brane is 
\begin{equation}
\label{induced1}
ds_4^2 = - h_n(r) dt^2 + \frac{1}{h_n(r)} dr^2 + 
r^2 (d \theta^2 + \sin^2 \theta d \varphi^2)
\end{equation}
where $h_n(r) = h_{n,+}(r) h_{n,-}(r)$ and 
$h_{n,\pm}(r) = 1 - (r_{\pm} / r)^{n+1}$. 

Using a separability condition 
$\Phi_{BR} = e^{-i \omega t} R_{\ell}(r) Y_{\ell, \tilde{m}}(\theta, \varphi)$,
it is easy to show that the wave equation $(\Box - m^2) \Phi_{BR} = 0$ reduces 
to the following radial equation
\begin{eqnarray}
\label{radial1}
& & x (x^{n+1} - x_{+}^{n+1})^2 (x^{n+1} - x_{-}^{n+1})^2 
\frac{d^2 R}{d x^2}
+ (x^{n+1} - x_{+}^{n+1}) (x^{n+1} - x_{-}^{n+1})
                                 \\   \nonumber
& & \hspace{0.5cm} \times
\left[ (n+1) x^{n+1} (2 x^{n+1} - x_{+}^{n+1} - x_{-}^{n+1}) - 2 n
       (x^{n+1} - x_{+}^{n+1}) (x^{n+1} - x_{-}^{n+1}) \right]
\frac{d R}{d x} 
                                 \\   \nonumber
& & \hspace{2.0cm}
+ \Bigg[ x^{4 n + 5} - \ell (\ell + 1) x^{2 n + 1}
         (x^{n+1} - x_{+}^{n+1}) (x^{n+1} - x_{-}^{n+1})
                                  \\   \nonumber
& & \hspace{4.0cm}
        + \frac{m^2}{\omega^2 v^2} x^{2n+3}
        \left\{ x^{n+1} (x_{+}^{n+1} + x_{-}^{n+1} ) - 
        x_{+}^{n+1} x_{-}^{n+1} \right\} \Bigg] R = 0
\end{eqnarray}
where $v = \sqrt{1 - m^2 / \omega^2}$, $x \equiv \omega v r$ and 
$x_{\pm} \equiv \omega v r_{\pm}$. From now on we take the massless 
case ($m=0$) for simplicity.

From Eq.(\ref{radial1}) it is straightforward to derive the following
Schr\"{o}dinger-like equation
\begin{equation}
\label{schro1}
-\frac{d^2 \psi}{d r_*^2} + V_{eff}^{BR} \psi = \omega^2 \psi
\end{equation}
where $\psi \equiv r R$ and the tortoise coordinate $r_*$ is 
\begin{equation}
\label{tortoise1}
r_{*} = \int \frac{d r}{h_{n}(r)}.
\end{equation}
The effective potential $V_{eff}^{BR}$ in Eq.(\ref{schro1}) is 
\begin{equation}
\label{effectivep1}
V_{eff}^{BR} = \frac{h_n(r)}{r^2}
\left[\ell (\ell + 1) + (n+1) \left\{h_{n,+}(r) + h_{n,-}(r) - 2 h_n(r) 
                                                       \right\} \right].
\end{equation}
Although it seems to be impossible to carry out the integration in
Eq.(\ref{tortoise1}) for the arbitrary $n$, we can easily infer the behavior
of the tortoise coordinate in the near-horizon and asymptotic regimes
\begin{equation}
\label{tortoise2}
\lim_{r \rightarrow r_+} r_* \sim \frac{r_+}{(n+1) h_{n,-}(r_+)}
\ln (r - r_+),
\hspace{2.0cm}
\lim_{r \rightarrow \infty} r_* \sim r.
\end{equation}
For $n=0$, $1$, and $2$ the explicit expressions of $r_*$ are 
\begin{eqnarray}
\label{tortoise3}
r_*&=& r + \frac{1}{r_+ - r_-}
\left[ r_+^2 \ln (r - r_+) - r_-^2 \ln (r - r_-) \right]
\hspace{3.8cm} \mbox{(n=0)}
                                            \\   \nonumber
r_*&=& r + \frac{1}{2 (r_+^2 - r_-^2)}
\left[r_+^3 \ln \frac{r - r_+}{r + r_+} - r_-^3 \ln \frac{r - r_-}{r + r_-}
                                               \right]
\hspace{3.6cm} \mbox{(n=1)}
                                             \\   \nonumber
r_*&=& r + \frac{1}{3 (r_+^3 - r_-^3)}
\left[ r_+^4 \ln \frac{r - r_+}{\sqrt{r^2 + r_+ r + r_+^2}}
     - r_-^4 \ln \frac{r - r_-}{\sqrt{r^2 + r_- r + r_-^2}}  \right]
                                             \\   \nonumber
& & \hspace{1.0cm}
    -\frac{1}{\sqrt{3} (r_+^3 - r_-^3)}
\left[ r_+^4 \tan^{-1} \frac{2r + r_+}{\sqrt{3} r_+} - 
      r_-^4 \tan^{-1} \frac{2r + r_-}{\sqrt{3} r_-} \right].
\hspace{1.0cm} \mbox{(n=2)}
\end{eqnarray}

\begin{figure}[ht!]
\begin{center}
\epsfysize=6.5 cm \epsfbox{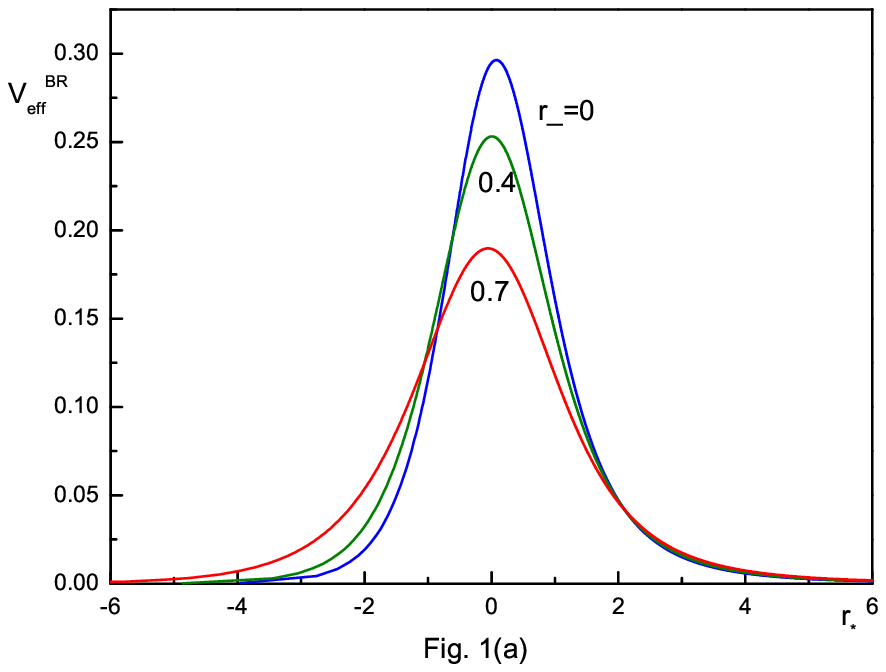}
\epsfysize=6.5 cm \epsfbox{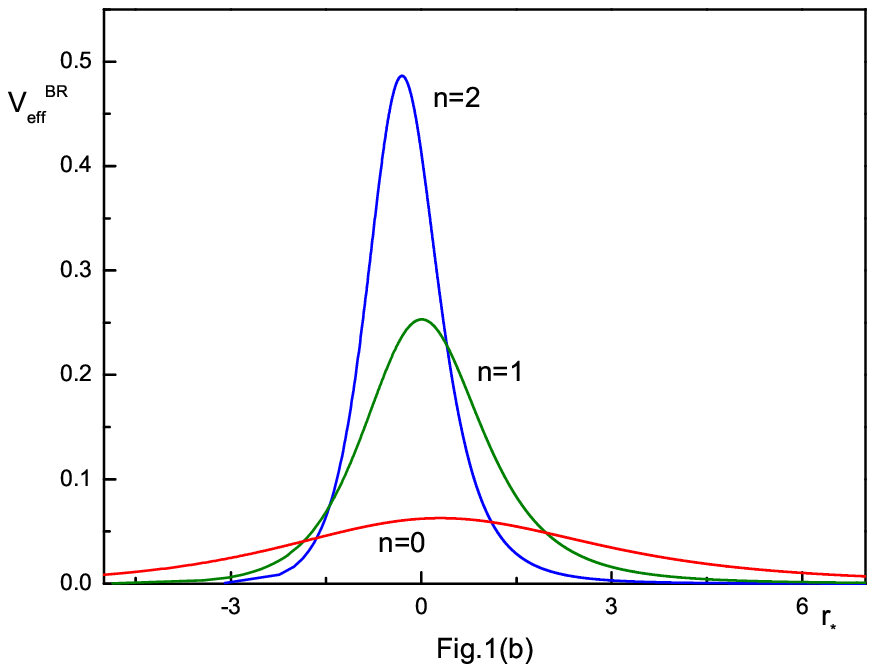}
\caption[fig1]{(a) Plot of the effective potential $V_{eff}^{BR}$ with respect to 
the tortoise coordinate $r_*$ for $r_- = 0$, $0.4$ and $0.7$ when $n=1$, $\ell =0$
and $r_+ = 1$. This figure indicates the absorption cross section enhances when
$r_-$ approaches to the extremal limit. This fact can be deduced also from the 
Hawking temperature. (b) Plot of the effective potential $V_{eff}^{BR}$ with 
respect to the tortoise coordinate $r_*$ for $n = 0$, $1$ and $2$ when 
$r_- = 0.4$, $r_+ = 1$ and $\ell = 0$. This figure indicates that the absorption
ability of black hole may be suppressed with increase of $n$. This can be 
understood from the monotonically decreasing behavior of the critical radius
$r_c$ with respect to $n$.}
\end{center}
\end{figure}

Fig. 1(a) is a plot of $V_{eff}^{BR}$ at $r_- = 0$, $0.4$, and $0.7$. This 
figure shows the barrier height becomes lower when the inner horizon
parameter $r_-$ increases. This implies that the absorptivity of the black hole should
enhance with increase of $r_-$. This fact can be conjectured by 
Hawking temperature in Eq.(\ref{mqst}). The Hawking temperature with fixed
$r_+$ has a maximum $(n+1) / 4 \pi r_+$ at the Schwarzschild limit and becomes
lower and lower with increase of $r_-$, and eventually goes to zero in the
extremal limit. Since the frozen matter in general absorbs something easily but
does not emit well to move to the equilibrium state,
one can understand Fig. 1(a) from the Hawking 
temperature. Fig. 1(b) is a plot of $V_{eff}^{BR}$  at $n=0$, $1$, and $2$
when $r_- = 0.4$. This figure shows the barrier height becomes higher when
the number of the extra dimensions increases, which indicates that the 
absorption cross section decreases with increase of $n$. This can be understood
from the fact that the critical radius $r_c \equiv \sqrt{\sigma_{\infty} / \pi}$,
where $\sigma_{\infty}$ is an high-energy limit of the total cross section, 
monotonically decreases with
increase of $n$\cite{emp00}.  

Another property of the radial equation (\ref{radial1}) is its real nature:
if $R$ is a solution of (\ref{radial1}), $R^*$ is a solution too. The Wronskian
of $R$ and $R^*$ can be evaluated from Eq.(\ref{radial1});
\begin{equation}
\label{wronskian1}
W[R^*, R]_x \equiv R^* \frac{d R}{d x} - R \frac{d R^*}{d x}
= \frac{C_n x^{2 n}}{(x^{n+1} - x_+^{n+1}) (x^{n+1} - x_-^{n+1})}
\end{equation}
where $C_n$ is a $n$-dependent integration constant.

Now, we would like to derive the solutions of the radial equation 
(\ref{radial1}). The solution which is convergent around the near-horizon
region can be expanded as 
\begin{equation}
\label{nhorizon1}
{\cal G}_{n,\ell} (x, x_+, x_-) = e^{\lambda_n \ln |x - x_+|}
\sum_{N=0}^{\infty} d_{\ell,N} (x - x_+)^N
\end{equation}
where $\lambda_n$ is a $n$-dependent pure imaginary quantity
\begin{equation}
\label{lambda1}
\lambda_n = -i \frac{x_+^{n+2}}{(n+1) (x_+^{n+1} - x_-^{n+1})}.
\end{equation}
When $n=0$, $d_{\ell,N}$ satisfies the following recursion relation
\begin{eqnarray}
\label{recursion1}
& & \hspace{2.0cm}
d_{\ell,N} \left[(x_+ - x_-)^2 (N + \lambda_0)^2 + x_+^4 \right]
                                                      \\   \nonumber
& &
+ d_{\ell,N-1} \left[(N + \lambda_0 - 1) (2 N + 2 \lambda_0 - 1) (x_+ - x_-)
                     + 4 x_+^3 - \ell (\ell + 1) (x_+ - x_-) \right]
                                                      \\   \nonumber
& & + d_{\ell,N-2} \left[(N + \lambda_0 - 1) (N + \lambda_0 - 2) + 6 x_+^2
                     - \ell (\ell + 1) \right]
+ 4 x_+ d_{\ell,N-3} + d_{\ell,N-4} = 0.
\end{eqnarray}
In appendix A we will present the recursion relation for $n=1$ case. 
Although similar
relations can be derived for $n \geq 2$, they are very lengthy and thus the 
explicit expressions will not be given in this paper.

The solutions of Eq.(\ref{radial1}) which are convergent in the asymptotic region
can be also expanded as
\begin{equation}
\label{asymp1}
{\cal F}_{n,\ell (\pm)}(x, x_+, x_-) = (\pm i)^{\ell + 1} e^{\mp i x}
(x - x_+)^{\pm \lambda_n} \sum_{N=0}^{\infty} \tau_{N (\pm)} x^{-(N+1)}
\end{equation}
where ${\cal F}_{n,\ell (+)}$ and ${\cal F}_{n,\ell (-)}$ are ingoing and outgoing
solutions respectively. It is worthwhile noting that ${\cal F}_{n,\ell (+)}$ is 
a complex conjugate of ${\cal F}_{n,\ell (-)}$. The coefficient $\tau_{N (-)}$
satisfies the following recursion relation when $n=0$:
\begin{eqnarray}
\label{recursion2}
& & \hspace{2.0cm}
[C_1 - 2 i (N+1)] \tau_{N (-)} + [N(N+1) - B_1 N + C_2] \tau_{N-1 (-)} 
                                                   \\   \nonumber
& &
+ [A_1 (N-1) N - B_2 (N-1) + C_3] \tau_{N-2 (-)}  
+ [A_2 (N-2) (N-1) - B_3 (N-2) + C_4] \tau_{N-3 (-)} 
                                                    \\  \nonumber
& & \hspace{1.0cm}
+ [A_3 (N-3) (N-2) - B_4 (N-3)] \tau_{N-4 (-)} 
+ A_4 (N-4) (N-3) \tau_{N-5 (-)}
= 0
\end{eqnarray}
where $\tau_{0 (-)} = 1$, $\tau_{N (+)} = \tau_{N (-)}^*$, and 
\begin{eqnarray}
\label{japda1}
& & A_1 = -2 (x_+ + x_-)
\hspace{3.0cm}
    A_2 = x_+^2 + 4 x_+ x_- + x_-^2
                                     \\    \nonumber
& & A_3 = -2 x_+ x_- (x_+ + x_-)
\hspace{2.0cm}
    A_4 = x_+^2 x_-^2
                                     \\    \nonumber
& & B_1 = 2 \left(1 - i \frac{x_+^2 - 2 x_-^2}{x_+ - x_-} \right)
\hspace{1.8cm}
    B_2 = -3 (x_+ + x_-) + 2 i \frac{x_- (x_+^2 - 3 x_+ x_- - x_-^2)}{x_+ - x_-}
                                     \\    \nonumber
& & B_3 = (x_+^2 + 4 x_+ x_- + x_-^2) + 2 i \frac{x_+ x_-^2(x_+ + 2 x_-)}
                                                 {x_+ - x_-}
\hspace{1.0cm}
B_4 = -x_+ x_- (x_+ + x_-) - 2 i \frac{x_+^2 x_-^3}{x_+ - x_-}
                                            \\    \nonumber
& &C_1 = -\frac{2 x_-^2}{x_+ - x_-} + 2 i
\hspace{1.5cm}
C_2 = \left[\frac{x_-^2 (2 x_+^2 - 2 x_+ x_- - x_-^2)}{(x_+ - x_-)^2}
               - \ell (\ell + 1) \right]
         - i \frac{2 x_+^2 - 3 x_-^2}{x_+ - x_-}
                                            \\    \nonumber
& &C_3 = \left[\frac{2 x_+ x_-^4}{(x_+ - x_-)^2} + \ell (\ell + 1) (x_+ + x_-)
         \right] + i \frac{x_- (2 x_+^2 - 3 x_+ x_- - x_-^2)}{x_+ - x_-}
                                            \\    \nonumber
& &C_4 = -\left[\frac{x_+^2 x_-^4}{(x_+ - x_-)^2} + \ell (\ell + 1) x_+ x_-
          \right] + i \frac{x_+ x_-^3}{x_+ - x_-}.
\end{eqnarray}
The recursion relation for $n=1$ case is explicitly given in appendix A.

Using Eq.(\ref{wronskian1}) it is easy to show
\begin{eqnarray}
\label{wronskian2}
& &W[{\cal G}_{n, \ell}^*, {\cal G}_{n, \ell}]_x \equiv 
{\cal G}_{n, \ell}^* \partial_x {\cal G}_{n, \ell} - {\cal G}_{n, \ell}
\partial_x {\cal G}_{n, \ell}^* = 
\frac{-2 i |g_{n,\ell}|^2 x_+^2 x^{2 n}}
     {(x^{n+1} - x_+^{n+1}) (x^{n+1} - x_-^{n+1})}
                                                       \\   \nonumber
& &W[{\cal F}_{n,\ell (+)}, {\cal F}_{n,\ell (-)}]_x \equiv
{\cal F}_{n,\ell (+)} \partial_x {\cal F}_{n,\ell (-)} - {\cal F}_{n,\ell (-)}
\partial_x {\cal F}_{n,\ell (+)} = 
\frac{2 i x^{2 n}}{(x^{n+1} - x_+^{n+1}) (x^{n+1} - x_-^{n+1})}
\end{eqnarray}
where $g_{n, \ell} \equiv d_{\ell, 0}$.

Now, we would like to show how the coefficient $g_{n, \ell}$ is related to the 
partial scattering amplitude. Since the real scattering solution, {\it say}
$R_{n,\ell}$, should be ingoing wave at the near-horizon region and mixture
of ingoing and outgoing waves at the asymptotic region, we can express it in the 
form
\begin{eqnarray}
\label{rsolution1}
& &R_{n,\ell}\stackrel{x \rightarrow x_+}{\sim} g_{n,\ell}(x-x_+)^{\lambda_n}
\left[1 + O(x -x_+)\right]   \\ \nonumber
& &R_{n,\ell}\stackrel{x \rightarrow \infty}{\sim} i^{\ell + 1} \frac{2 \ell +1}
{2 x}\left[e^{-ix + \lambda_n \ln|x-x_+|} - (-1)^{\ell}
S_{n,\ell}(x_+, x_-) e^{ix - \lambda_n\ln|x - x_+|}\right]
+ O\left(\frac{1}{x^2}\right)
\end{eqnarray}
where $S_{n,\ell}$ is a partial scattering amplitude. If we define a phase shift
$\delta_{n,\ell}$ as $S_{n,\ell} \equiv e^{2 i \delta_{n,\ell}}$, the second equation
of Eq.(\ref{rsolution1}) can be written as
\begin{equation}
\label{rinfty}
R_{n,\ell}\stackrel{x \rightarrow \infty}{\sim} \frac{2 \ell +1}{x} e^{i \delta_
{n,\ell}} \sin \left[x + i \lambda_n \ln|x-x_+|- \frac{\pi}{2}\ell + \delta_{n,\ell}
\right] + O\left(\frac{1}{x^2}\right).
\end{equation}

Next let us consider the Wronskian 
$W[R_{n,\ell}^*, R_{n,\ell}]_x \equiv 
R_{n,\ell}^{\ast} \partial_x R_{n,\ell}-R_{n,\ell} \partial_x R_{n,\ell}^{\ast}$. 
From the first equation of Eq.(\ref{rsolution1}) it is straightforward to show
$W[R_{n,\ell}^*, R_{n,\ell}]_x = W[{\cal G}_{n,\ell}^*, {\cal G}_{n,\ell}]_x$
while Eq.(\ref{rinfty}) makes the Wronskian in the form
\begin{equation}
\label{wronskian3}
W[R_{n,\ell}^*, R_{n,\ell}]_x = 
\frac{-i (2 \ell + 1)^2 x^{2 n}}{(x^{n+1} - x_+^{n+1}) (x^{n+1} - x_-^{n+1})}
e^{-2 \beta_{n,\ell}} \sinh 2 \beta_{n, \ell}
\end{equation}
where $\delta_{n, \ell}$ is assumed as $\delta_{n, \ell} \equiv \eta_{n, \ell}
+ i \beta_{n, \ell}$. Therefore, equating those two Wronskians yield a relation
\begin{equation}
\label{gell}
|g_{n,\ell}|^2 = \frac{\left( \ell +\frac{1}{2}\right)^2}{x_+^2}
\left(1-e^{-4
\beta_{n,\ell}}\right).
\end{equation}
Since $1-e^{-4 \beta_{n,\ell}} = 1 - |S_{n, \ell}(x_+,x_-)|^2$ is a transmission 
coefficient\footnote{This can be proven directly by computing the ingoing flux
$j_{in}$ and outgoing flux $j_{out}$. Then it is easy to show 
$|S_{n, \ell}|^2 = |j_{out} / j_{in}|$, which completes the proof.}, we can compute
the absorption cross section using Eq.(\ref{partialabs}) if we know $g_{n, \ell}$.

Next we would like to discuss how to compute $g_{n, \ell}$ by matching the 
asymptotic solution (\ref{asymp1}) with the near-horizon solution (\ref{nhorizon1}).
In order to discuss it properly it is convenient\cite{sanc78,jung04-2} to 
introduce a new wave 
solution $\tilde{R}_{n,\ell}(x, x_+, x_-)$, which differs from 
$R_{n,\ell}(x, x_+, x_-)$ in its normalization. It is normalized in such a way that
\begin{equation}
\label{tilda1}
\tilde{R}_{n,\ell}(x, x_+, x_-)\stackrel{x \rightarrow x_+}{\sim}
(x - x_+)^{\lambda_n}\left[ 1 + O (x - x_+)\right].
\end{equation}
Since ${\cal F}_{n,\ell (\pm)}$ in Eq.(\ref{asymp1}) are two linearly independent
solutions of the radial equation (\ref{radial1}), one may write 
$\tilde{R}_{n,\ell}$ as a combination of them
\begin{equation}
\label{tilda2}
\tilde{R}_{n,\ell}(x, x_+, x_-) = f_{n,\ell}^{(-)}(x_+, x_-) 
{\cal F}_{n,\ell(+)} (x, x_+, x_-) +
f_{n, \ell}^{(+)}(x_+, x_-){\cal F}_{n,\ell(-)}(x, x_+, x_-)
\end{equation}
where the coefficients $f_{n,\ell}^{(\pm)}(x_+, x_-)$ are called the jost functions.
Using Eq.(\ref{wronskian2}) one can compute $f_{n,\ell}^{(\pm)}$ as following:
\begin{eqnarray}
\label{jost1}
f_{n,\ell}^{(\pm)}(x_+, x_-) &=& \pm \frac{(x^{n+1}-x_+^{n+1})(x^{n+1}-x_-^{n+1})}
{2i x^{2n}} W[{\cal F}_{n,\ell (\pm)}, \tilde R]_x
                                                       \\   \nonumber
&=& \pm \frac{\omega (r^{n+1} - r_+^{n+1}) (r^{n+1} - r_-^{n+1})}{2i r^{2n}}
     W[{\cal F}_{n,\ell (\pm)}, \tilde R]_r.
\end{eqnarray}
Inserting the explicit form of ${\cal F}_{n,\ell (\pm)}$ into Eq. (\ref{tilda2})
and comparing it with the asymptotic expression of $R_{n, \ell}$ in 
Eq. (\ref{rsolution1}), one can easily derive the following two relations
\begin{eqnarray}
\label{scattering1}
S_{n, \ell} (x_+, x_-)&=&\frac {f_{n, \ell}^{(+)}(x_+, x_-)}  
                               {f_{n, \ell}^{(-)}(x_+, x_-)} 
                                              \\  \nonumber
f_{n, \ell}^{(-)}(x_+, x_-)&=&\frac{\ell + \frac{1}{2}} {g_{n, \ell} (x_+, x_-)}.
\end{eqnarray}
Thus once we know the jost functions, we can compute the scattering quantities 
such as $S_{n, \ell} (x_+, x_-)$ and $g_{n, \ell} (x_+, x_-)$.
Combining Eq.(\ref{gell}) and (\ref{scattering1}), we can express the greybody 
factor in terms of the jost function
\begin{equation}
\label{jost2}
1 - |S_{n,\ell}(x_+, x_-)|^2 = 
\frac{x_+^2}{|f_{n,\ell}^{(-)} (x_+, x_-)|^2}.
\end{equation}

Since the relation between the absorption cross section $\sigma_{n,\ell}^{BR}$ and
the greybody factor for the brane-localized scalar is given in Eq.(\ref{partialabs})
with fixing $D=4$, it is easy to show
\begin{equation}
\label{brsection2}
\sigma_{n,\ell}^{BR} = \frac{\pi}{\omega^2} (2 \ell + 1)
\left[1 - |S_{n,\ell}(x_+, x_-)|^2 \right]
= \frac{\pi (2 \ell + 1) r_+^2}{|f_{n,\ell}^{(-)} (x_+, x_-)|^2}.
\end{equation}
Thus one can compute the absorption cross section from the jost function
$f_{n,\ell}^{(-)} (x_+, x_-)$. In next section we will compute the jost functions
numerically by applying the analytic continuation.

\section{Absorption and Emission for the brane-localized scalar}
In this section we will compute the jost functions $f_{n,\ell}^{(\pm)} (x_+, x_-)$
numerically. The computational procedure is as following. Firstly, we note that 
$\tilde{R}_{n,\ell}(x, x_+, x_-)$ defined in Eq.(\ref{tilda1}) can be obtained 
from ${\cal G}_{n,\ell}(x, x_+, x_-)$ by $\tilde{R}_{n,\ell}(x, x_+, x_-)
= {\cal G}_{n,\ell}(x, x_+, x_-)|_{d_{\ell, 0} =1}$. This can be understood from
a convergent expansion
around the near-horizon region. The other expression (\ref{tilda2}) is of course
a convergent expansion around the asymptotic region. Since, however, the domains
of convergence for those two expressions are different, we cannot use them directly
for the computation of the jost functions. In other words, we need two expressions
which have common domain of convergence. This is achieved by analytic 
continuation. The solution which is a power series in the neighborhood of an 
arbitrary point $x = b$ can be straightforwardly obtained from the radial 
equation (\ref{radial1}), whose formal form is 
\begin{equation}
\label{contin1}
\varphi_{n,\ell}(x, x_+, x_-) = (x - x_+)^{\lambda_n}
\sum_{N=0}^{\infty} D_N (x - b)^N.
\end{equation}
For $n=0$ the coefficient $D_N$ satisfies the following 
recursion relation:
\begin{eqnarray}
\label{contin2}
& &\tilde{A}_4 (N+5) (N+6) D_{N+6} + 
\left[\tilde{A}_3 (N+4) (N+5) + \tilde{B}_3 (N+5) \right] D_{N+5}
                                                            \\   \nonumber
& & \hspace{2.0cm}
  + \left[\tilde{A}_2 (N+3) (N+4) + \tilde{B}_2 (N+4) + \tilde{C}_4\right] D_{N+4}
                                                            \\  \nonumber
& & \hspace{2.0cm}
+ \left[\tilde{A}_1 (N+2) (N+3) + \tilde{B}_1 (N+3) + \tilde{C}_3 \right] D_{N+3}
                                                            \\   \nonumber
& &+ \left[(N+1) (N+2) + 2 (\lambda_0 + 1) (N+2) + \tilde{C}_2 \right] D_{N+2}
+ \tilde{C}_1 D_{N+1} + D_N = 0
\end{eqnarray}
where
\begin{eqnarray}
\label{japda2}
& &\tilde{A}_1 = 2 \left[(b - x_+) + (b - x_-) \right]
\hspace{1.0cm}
    \tilde{A}_2 = (b - x_+)^2 + 4 (b - x_+) (b - x_-) + (b - x_-)^2
                                                           \\   \nonumber
& &\tilde{A}_3 = (b - x_+)^2 (b - x_-)^2
\hspace{1.0cm}
   \tilde{B}_1 = (2 \lambda_0 + 3) (b - x_+) + (4 \lambda_0 + 3) (b - x_-)
                                                            \\    \nonumber
& &\tilde{B}_2 = (b - x_+)^2 + 4 (\lambda_0 + 1) (b - x_+) (b - x_-) + 
                 (2 \lambda_0 + 1) (b - x_-)^2
                                                           \\   \nonumber
& &\tilde{B}_3 = (b - x_+) (b - x_-) \left[(b - x_+) + (2 \lambda_0 + 1) (b - x_-)
                                              \right]
\hspace{1.0cm}
\tilde{C}_1 = 4 b
                                                           \\    \nonumber
& &\tilde{C}_2 = \lambda_0 (\lambda_0 + 1) + 6 b^2 - \ell (\ell + 1)
                                                           \\    \nonumber
& &\tilde{C}_3 = \left[\lambda_0 - \ell (\ell + 1) \right] (b - x_+)
                + \left[ \lambda_0 (2 \lambda_0 + 1) - \ell (\ell + 1)
                         \right] (b - x_-) + 4 b^3
                                                            \\    \nonumber
& &\tilde{C}_4 = \lambda_0^2 (b - x_-)^2 + \left[\lambda_0 - \ell (\ell + 1) \right]
                 (b - x_+) (b - x_-) + b^4.
\end{eqnarray}
The recursion relation for $n=1$ is explicitly given in appendix A. The 
recursion relation for each $n$ enables us to compute all 
the $D_N$'s ($N \geq 2$) in terms
of $D_0$ and $D_1$. Since $D_0$ and $D_1$ are expressed in terms of 
$\varphi_{n,\ell}(b)$ and $\partial_x \varphi_{n,\ell}(b)$, the analytic 
continuation can be directly achieved. In actual computer calculation the asymptotic
region is identified by $r \sim 1000$ and the analytic continuation procedure is 
repeated over and over to make a common domain of convergence.

\begin{figure}[ht!]
\begin{center}
\epsfysize=6.5 cm \epsfbox{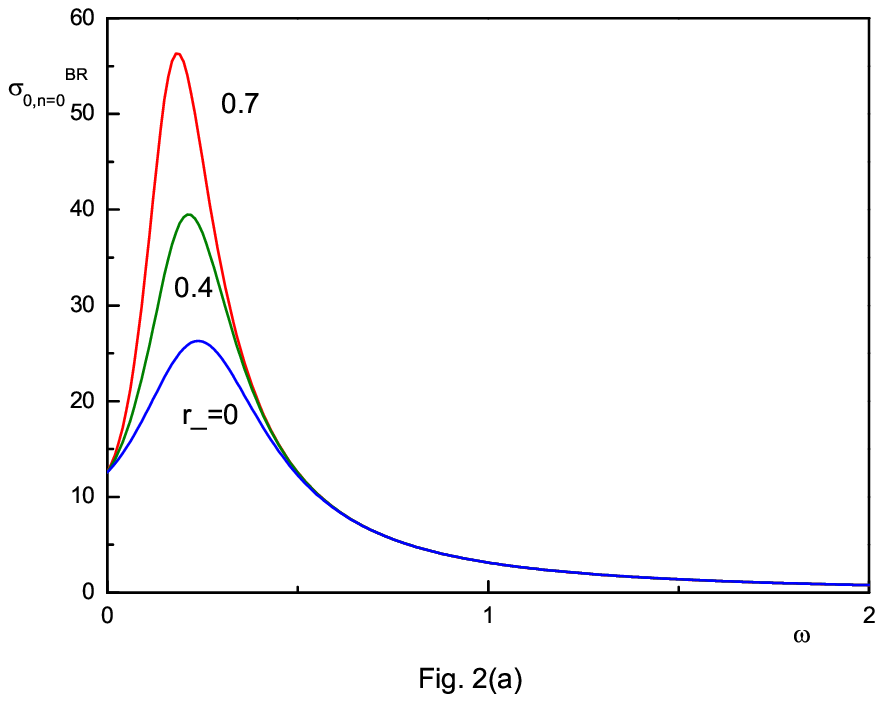}
\epsfysize=6.5 cm \epsfbox{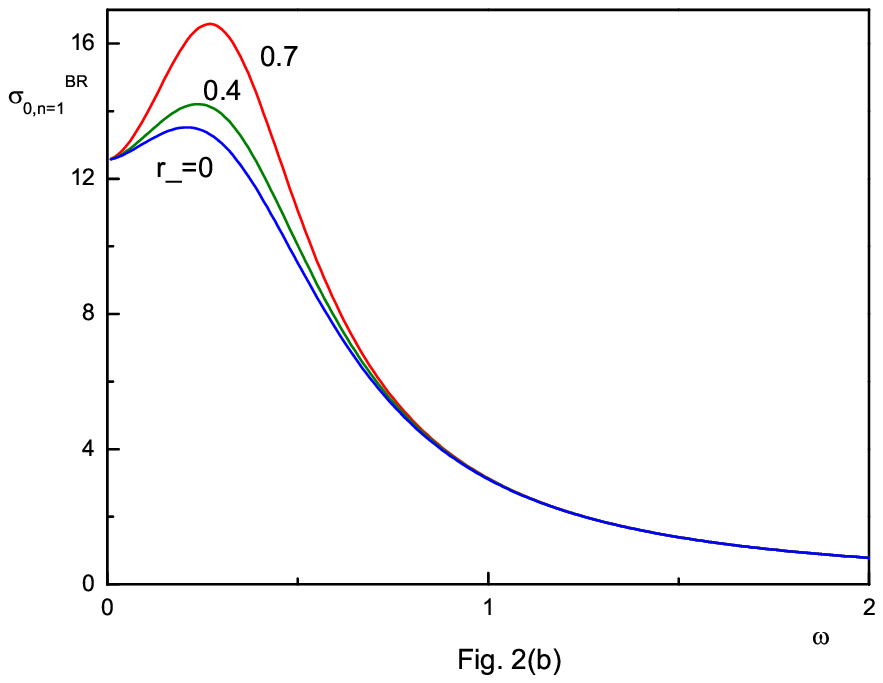}
\epsfysize=6.5 cm \epsfbox{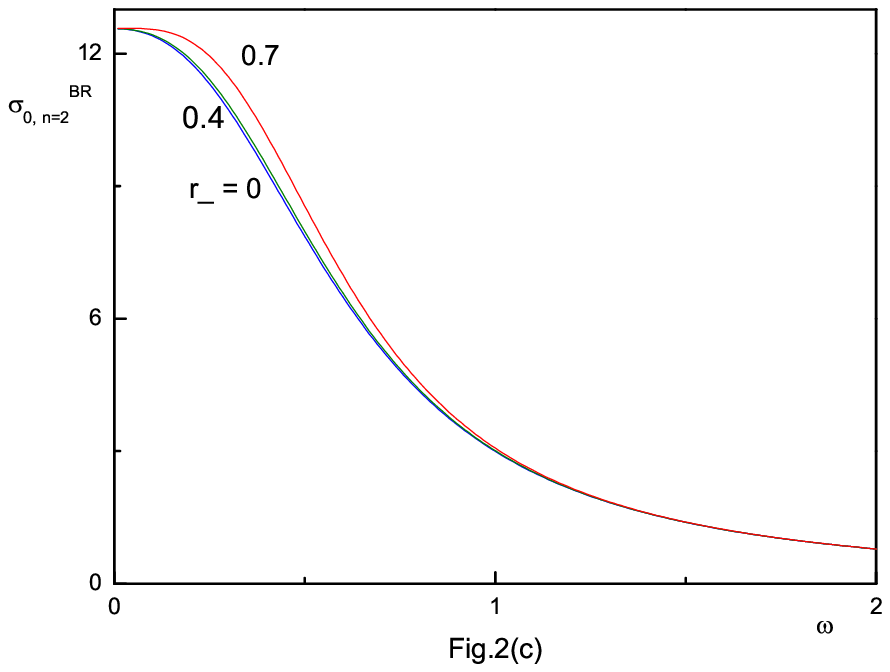}
\caption[fig2]{Plot of the partial absorption cross section for 
s-wave ($\ell = 0$) when
$r_+ = 1$ and $n=0$ (Fig. 2(a)), $r_+ = 1$ and $n=1$ (Fig. 2(b)), and 
$r_+ = 1$ and $n=2$ (Fig. 2(c)) with varying $r_-$. These figures show how the 
absorption cross section is enhanced with increasing $r_-$. Regardless of $n$ the 
low-energy absorption cross section equals to the horizon area $4 \pi r_+^2$, 
which is an 
universal properties for the asymptotically flat and spherically symmetric
black holes}
\end{center}
\end{figure}

Fig. 2 shows $r_-$-dependence of the partial absorption cross section for 
s-wave ($\ell = 0$) when $n=0$, $1$ and $2$. As expected from the effective 
potential the absorption cross section is enhanced with increasing $r_-$. 
The sharp peaks in Fig. 2(a) ($n=0$ case) disappear in 
Fig. 2(c) ($n=2$ case). This seems to be because the presence of the extra dimensions 
strongly suppresses the absorptivity of the black hole, which is clear from
Fig. 1(b) or Fig. 3. 
Regardless of $n$ the low-energy absorption cross section equals to
$4 \pi r_+^2$, which is an universal property for the asymptotically flat and 
spherically symmetric black holes\cite{das97}. 

\begin{figure}[ht!]
\begin{center}
\epsfysize=6.3 cm \epsfbox{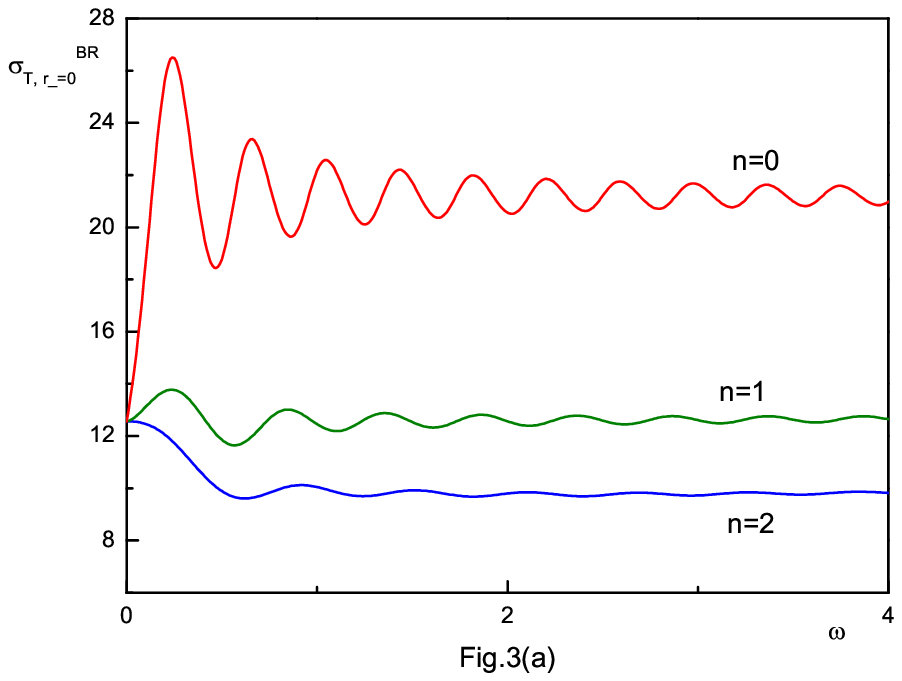}
\epsfysize=6.3 cm \epsfbox{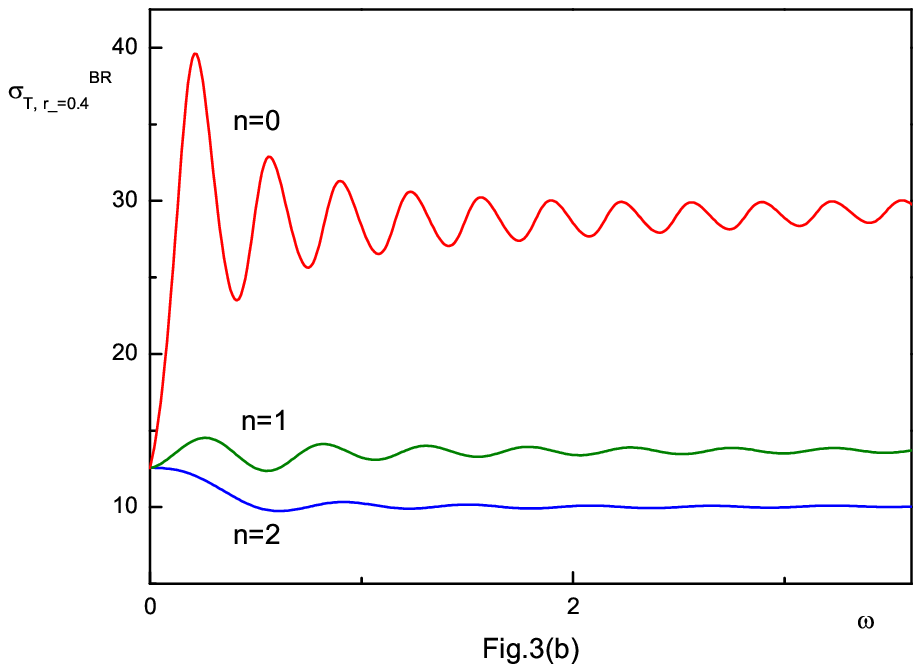}
\epsfysize=6.3 cm \epsfbox{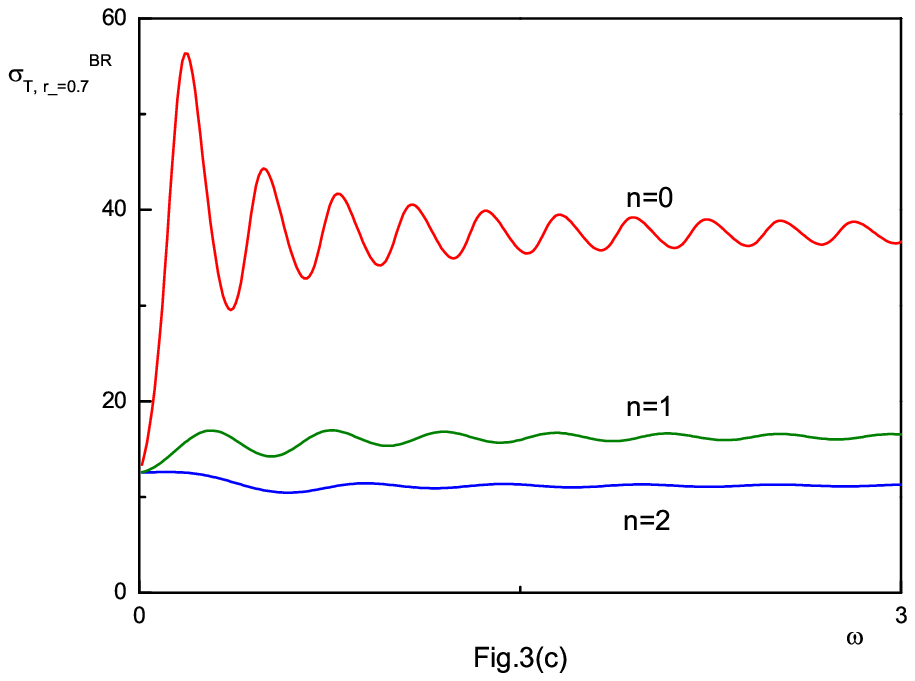}
\caption[fig3]{Plot of the total absorption cross section when $r_+ = 1$, and 
$r_- = 0$ (Fig. 3(a)), $0.4$ (Fig. 3(b)), and 0.7 (Fig. 3(c)) with varying $n$. 
These figures show how the existence of the extra dimensions suppresses the 
absorption cross section. Regardless of $n$ and/or $r_-$ the low energy absorption
cross section equals to the horizon area. This fact indicates that the higher
partial waves ($\ell \geq 1$) have vanishing low-energy limits.}  
\end{center}
\end{figure} 

Fig. 3 shows the $n$-dependence of the total absorption cross section $\sigma_T^{BR}$
when $r_- = 0$, $0.4$ and $0.7$. The oscillatory behavior of $\sigma_T^{BR}$ at 
$n=0$ indicates that each partial absorption cross section with different 
angular momentum $\ell$ has a peak in the different value of $\omega$. This
oscillatory behavior, however, disappears with increase of $n$. This is because 
the presence of the extra dimensions removes the peak of the each partial absorption
cross section as Fig. 2(c) exhibits. The fact that 
$\sigma_T^{BR}(\omega = 0) = 4 \pi r_+^2$
indicates that the partial absorption cross sections for $\ell \geq 1$ vanish
in the low-energy limit, which is in good agreement with Starobinsky's
formula\cite{sta73}.

From the Hawking formula (\ref{hawking1}) the brane emission, {\it i.e.} the energy 
emitted per unit time and energy interval $d \omega$, is given by
\begin{equation}
\label{hawking2}
\Gamma_D^{BR} = 
\frac{\omega^3 \sigma_{abs}}{2 \pi^2 \left(e^{\omega / T_H} - 1 \right)} d\omega
\end{equation}  
for the massless scalar localized on the brane.
The effect of the extra dimensions is included in the absorption cross section
$\sigma_{abs}$ and the Hawking temperature $T_H$. Since $\sigma_{abs}$ decreases and
$T_H$ increases in the presence of the extra dimensions, the complete emission 
spectrum is determined by the competition between the greybody and Planck
factors. Since, in addition, the inner horizon parameter $r_-$ generally increases
$\sigma_{abs}$ and decreases $T_H$, the emission rate in the presence of nonzero
$r_-$ is also determined by the competition between these two factor. 

Since, in general, the effect of the Planck factor is dominant compared to the 
greybody factor in the emission problem, we expect that the presence of the extra
dimensions enhances the emission spectrum. We also expect that 
contrary to the extra dimensions the presence of nonzero $r_-$ may decrease the 
emission spectrum of the black hole.

\begin{figure}[ht!]
\begin{center}
\epsfysize=6.5 cm \epsfbox{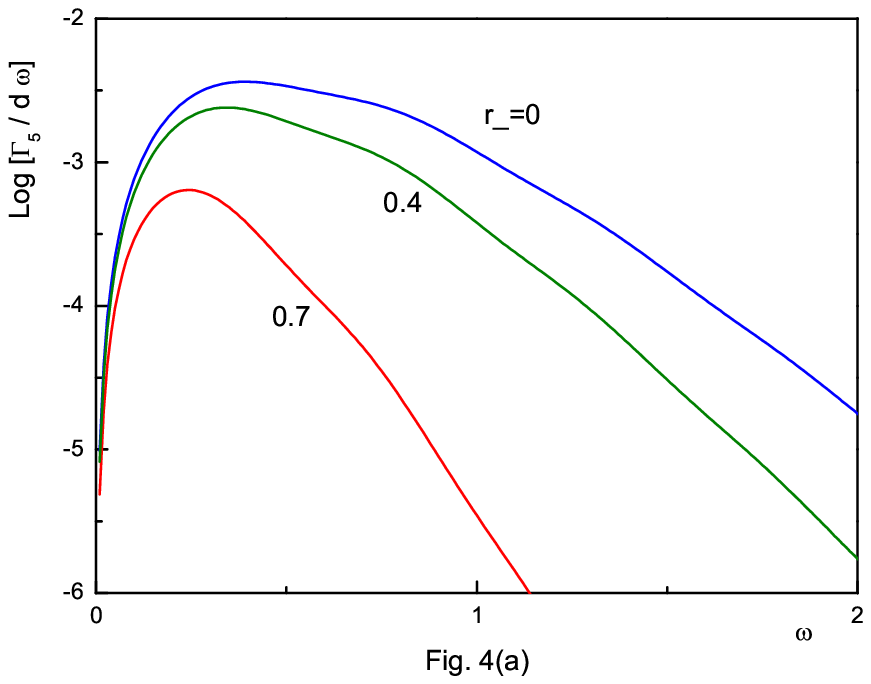}
\epsfysize=6.3 cm \epsfbox{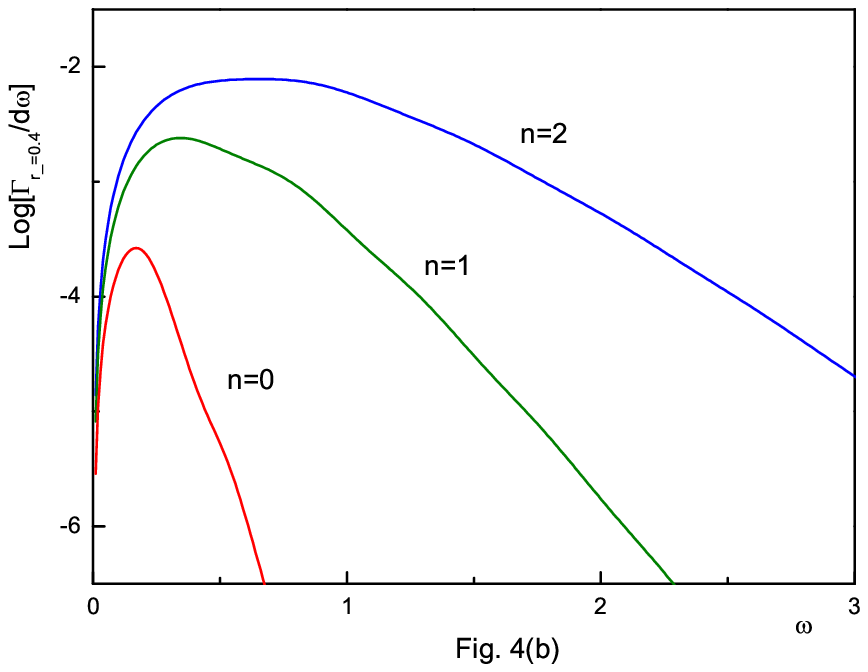}
\caption[fig4]{Log-plot of the emission rate when $r_+=1$ and $n=1$ with varying
$r_-$ (Fig. 4(a)), and when $r_+ = 1$ and $r_- = 0.4$ with varying $n$ (Fig. 4(b)).
These figures shows the emission rate reduces with increasing $r_-$. However, the
presence of the extra dimensions in general enhances the emission power. These
facts indicate that the Planck factor is dominant compared to the greybody
factor in the emission problem.}
\end{center}
\end{figure}

Fig. 4(a) is a $r_-$-dependence of the emission rate when $n=1$. As expected, the 
emission rate decreases with increase of $r_-$. As commented earlier the frozen
black hole seems to have a poor ability in the emission. Contrary to the Wien's type of
displacement of the peak, the peak moves to the opposite direction with increasing
$r_-$ in this figure. 
Fig. 4(b) is a $n$-dependence of the emission rate when $r_- = 0.4$. As 
expected again, the existence of the extra dimensions enhances the emission rate.

\section{wave equation for the bulk scalar}
The scalar equation $(\Box - m^2) \Phi_{BL} = 0$ for the bulk scalar 
in the background of the 
spacetime (\ref{space1}) reduces to the following radial equation
\begin{eqnarray}
\label{blradial1}
& & x (x^{n+1} - x_{+}^{n+1})^2 (x^{n+1} - x_{-}^{n+1})^2 
\frac{d^2 R}{d x^2}
+ (x^{n+1} - x_{+}^{n+1}) (x^{n+1} - x_{-}^{n+1})
                                 \\   \nonumber
& & \hspace{0.5cm} \times
\left[ (n+1) x^{n+1} (2 x^{n+1} - x_{+}^{n+1} - x_{-}^{n+1}) -  n
       (x^{n+1} - x_{+}^{n+1}) (x^{n+1} - x_{-}^{n+1}) \right]
\frac{d R}{d x} 
                                 \\   \nonumber
& & \hspace{2.0cm}
+ \Bigg[ x^{4 n + 5} - \ell (\ell + n +1) x^{2 n + 1}
         (x^{n+1} - x_{+}^{n+1}) (x^{n+1} - x_{-}^{n+1})
                                  \\   \nonumber
& & \hspace{4.0cm}
        + \frac{m^2}{\omega^2 v^2} x^{2n+3}
        \left\{ x^{n+1} (x_{+}^{n+1} + x_{-}^{n+1} ) - 
        x_{+}^{n+1} x_{-}^{n+1} \right\} \Bigg] R = 0
\end{eqnarray}
where $x \equiv \omega v r$ and $x_{\pm} \equiv \omega v r_{\pm}$. 
Of course, we assumed 
the separability condition $\Phi_{BL} = e^{-i \omega t} R(r) \tilde{Y}(\Omega)$, 
where $\tilde{Y}$ is an higher-dimensional spherical harmonics. From now on we take 
the massless case ($m=0$) for simplicity.

From Eq.(\ref{blradial1}) it is straightforward to derive the following 
Schr\"{o}dinger-like equation
\begin{equation}
\label{bl-schro1}
-\frac{d^2 \psi}{d r_{*}^2} + V_{eff}^{BL} \psi = \omega^2 \psi
\end{equation}
where the tortoise coordinate $r_*$ is given in Eq.(\ref{tortoise1}) and 
$\psi = r^{(n+2) / 2} R$. The effective potential $V_{eff}^{BL}$ is given by 
\begin{equation}
\label{bl-effective}
V_{eff}^{BL} = \frac{h_n(r)}{r^2}
\left[ \ell (\ell + n + 1) + \frac{n+2}{2}
      \left\{(n+1) (h_{n,+}(r) + h_{n,-}(r)) - \frac{3 n + 4}{2} h_n (r) \right\} 
                                                                         \right].
\end{equation}

\begin{figure}[ht!]
\begin{center}
\epsfysize=6.3 cm \epsfbox{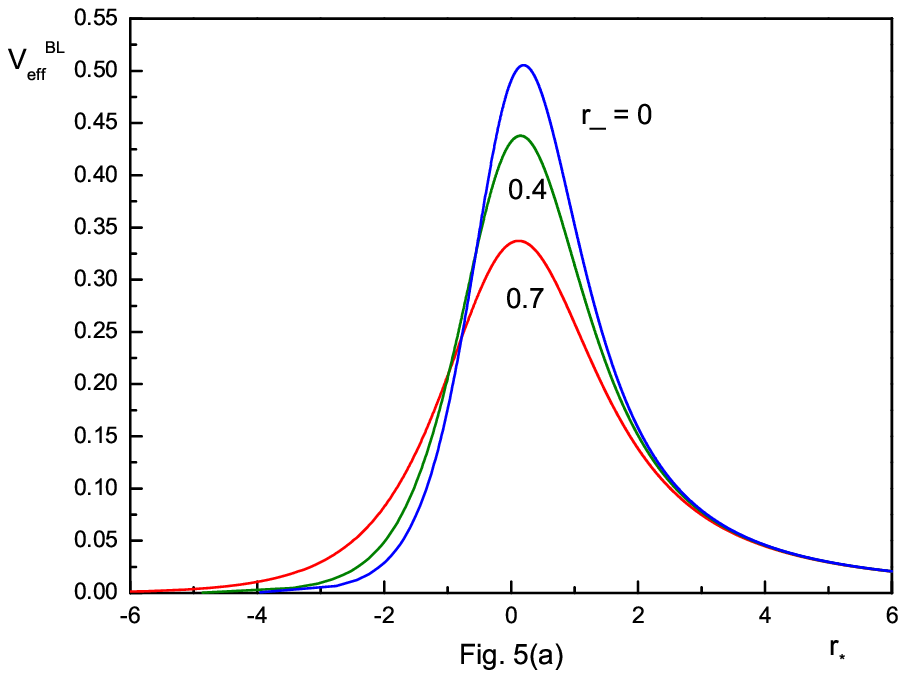}
\epsfysize=6.3 cm \epsfbox{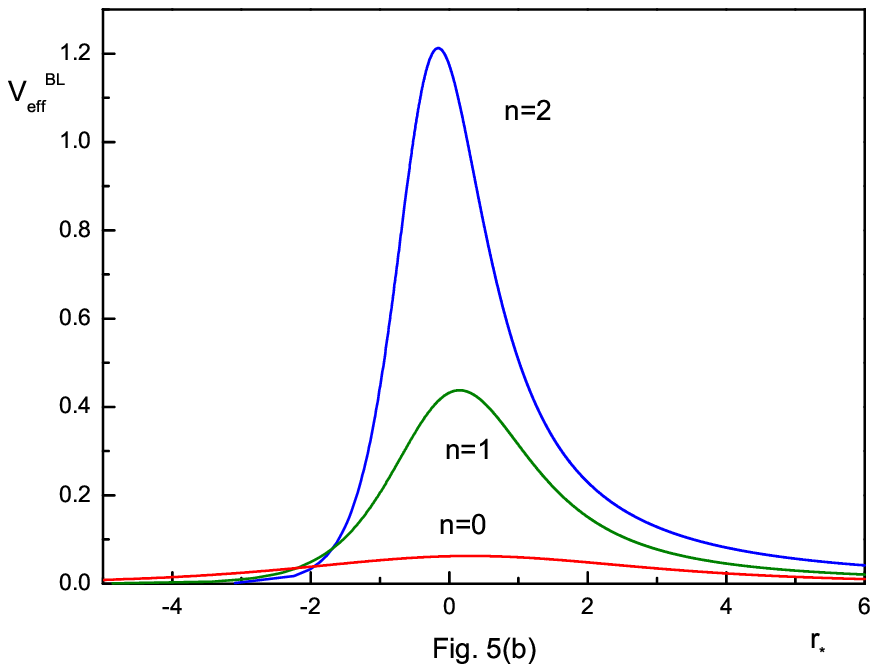}
\caption[fig5]{(a) Plot of the effective potential $V_{eff}^{BL}$ with respect to 
the tortoise coordinate $r_*$ for $r_- = 0$, $0.4$ and $0.7$ when $n=1$, $\ell =0$
and $r_+ = 1$. This figure indicates the absorption cross section enhances 
with increase of $r_-$. This fact can be deduced also from the 
Hawking temperature. (b) Plot of the effective potential $V_{eff}^{BL}$ with 
respect to the tortoise coordinate $r_*$ for $n = 0$, $1$ and $2$ when 
$r_- = 0.4$, $r_+ = 1$ and $\ell = 0$. This figure indicates that the absorptivity
of black hole for the bulk scalar may be suppressed with increase of $n$.} 
\end{center}
\end{figure}

Fig. 5 is a plot of $V_{eff}^{BL}$ with respect to the tortoise coordinate $r_{*}$.
Since the $r_{*}$-dependence of $V_{eff}^{BL}$ is very similar to the 
$r_{*}$-dependence of $V_{eff}^{BR}$ in Fig. 1, the absorption spectrum seems to 
be enhanced with increasing $r_-$. 
Also Fig. 5(b) indicates that the absorptivity for the bulk scalar by a black 
hole
may be suppressed with increase of $n$. As will be shown in the following section,
however, the disappearance of the oscillatory pattern in the total absorption
cross section occured in the case of the brane-localized scalar does not
happen in this case although the amplitude of oscillation decreases in the 
presence of the extra dimensions.

The calculational procedure for the numerical computation of the absorption 
cross section for the bulk scalar in the full-range of energy is similar to 
that for the case of the brane-localized scalar. The Wronskian of $R$ and $R^*$
which are solutions of the radial equation (\ref{blradial1}) is
\begin{equation}
\label{blwronskian1}
W[R^*, R]_x \equiv R^* \frac{d R}{d x} - R \frac{d R^*}{d x}
= \frac{C_n^{\prime} x^{n}}{(x^{n+1} - x_+^{n+1}) (x^{n+1} - x_-^{n+1})}
\end{equation}
where $C_n^{\prime} $ is a $n$-dependent integration constant.
Comparing Eq.(\ref{blwronskian1}) with the corresponding equation (\ref{wronskian1})
in the brane case, the only difference of the Wronskian is  the power of $x$ in the
numerator. However, this slight difference is crucial and makes completely
different absorption and emission spectra from those for the case of the 
brane-localized scalar.

The solutions of the radial equation (\ref{blradial1}) which has a convergent
range around the near-horizon region can be derived in the similar way to the 
brane-localized case
\begin{equation}
\label{blnhorizon1}
{\cal G}_{n,\ell} (x, x_+, x_-) = e^{\lambda_n \ln |x - x_+|}
\sum_{N=0}^{\infty} d_{\ell,N} (x - x_+)^N
\end{equation}
where $\lambda_n$ is given in Eq.(\ref{lambda1}).
For $n=0$ the recursion relation which the coefficient $d_{\ell,N}$ obeys should
be same with Eq.(\ref{recursion1}) because there is no distinction between the cases 
of bulk and brane. However, the recursion relations for $n \geq 1$ should be 
different from those in the brane-localized case. In appendix B the recursion
relation of $d_{\ell,N}$ for $n=1$ is explicitly given.
As in the case of the brane-localized scalar the recursion relations for $n \geq 2$ 
are 
not presented in the paper because they are too lengthy.

The solutions of Eq.(\ref{blradial1}) which are convergent in the asymptotic
region can be expressed as\cite{gub97-1}
\begin{equation}
\label{blasymp1}
{\cal F}_{n,\ell (\pm)}(x, x_+, x_-) = (\pm i)^{\ell + 1 + \frac{n}{2}}
\frac{e^{\mp ix}(x-x_+)^{\pm \lambda_n}}{x^{\frac{n}{2}}} 
\sum_{N=0}^{\infty} \tau_{N (\pm)} x^{-(N+1)}
\end{equation}
where ${\cal F}_{n,\ell (+)}$ and ${\cal F}_{n,\ell (-)}$ are ingoing and outgoing 
solutions respectively. 
The difference of Eq.(\ref{blasymp1}) from Eq.(\ref{asymp1}) is a power of 
$x$ in the denominator, which is also crucial in the following calculation.
As remaked just before, the recursion relation of 
$\tau_{N(\pm)}$ for $n=0$ should be same with Eq.(\ref{recursion2}). The recursion
relation for $n=1$ is explicitly given in appendix B.
Using Eq.(\ref{blwronskian1}) it is easy to show
\begin{eqnarray}
\label{blwronskian2}
& &W[{\cal G}_{n, \ell}^*, {\cal G}_{n, \ell}]_x \equiv 
{\cal G}_{n, \ell}^* \partial_x {\cal G}_{n, \ell} - {\cal G}_{n, \ell}
\partial_x {\cal G}_{n, \ell}^* = 
\frac{-2 i |g_{n,\ell}|^2 x_+^{n+2} x^{n}}
     {(x^{n+1} - x_+^{n+1}) (x^{n+1} - x_-^{n+1})}
                                                       \\   \nonumber
& &W[{\cal F}_{n,\ell (+)}, {\cal F}_{n,\ell (-)}]_x \equiv
{\cal F}_{n,\ell (+)} \partial_x {\cal F}_{n,\ell (-)} - {\cal F}_{n,\ell (-)}
\partial_x {\cal F}_{n,\ell (+)} =\frac{2 i x^{n}}
                                  {(x^{n+1} - x_+^{n+1}) (x^{n+1} - x_-^{n+1})}
\end{eqnarray}
where $g_{n, \ell} \equiv d_{\ell, 0}$.

The equation for the bulk scalar case corresponding to Eq(\ref{rsolution1})
for the brane-localized case is\cite{gub97-1}
\begin{eqnarray}
\label{blrsolution1}
& &R_{n,\ell}\stackrel{x \rightarrow x_+}{\sim} g_{n,\ell}(x-x_+)^{\lambda_n}
\left[1 + O(x -x_+)\right]   \\ \nonumber
& &R_{n,\ell}\stackrel{x \rightarrow \infty}{\sim} \frac{i^{\ell+1+\frac{n}{2}}
2^{\frac{n}{2}-1}}{\sqrt{\pi}x^{1+\frac{n}{2}}} \Gamma{(\frac{1+n}{2})}
\frac{(2 \ell +1+n)(\ell+n)!}{\ell ! n!}
                                                       \\   \nonumber    
& & \hspace{1.8cm} \times \bigg[e^{-ix+ \lambda_n \ln|x-x_+|} - (-1)^{\ell+\frac{n}{2}}
S_{n,\ell}(x_+,x_-) e^{ix+ \lambda_n \ln|x-x_+|}\bigg]
+ O\left(\frac{1}{x^{2+\frac{n}{2}}}\right)   
\end{eqnarray}
where $R_{n,\ell}$ is a real scattering solution of Eq.(\ref{blradial1}) and 
$S_{n,\ell}$ is a partial scattering amplitude. If we define a phase shift
$\delta_{n,\ell}$ as $S_{n,\ell} \equiv e^{2i \delta_{n,\ell}}$, the second 
equation of Eq.(\ref{blrsolution1}) can be written as
\begin{eqnarray}
\label{blrinfty}
R_{n,\ell}\stackrel{x \rightarrow \infty}{\sim} 
& & \frac{2^{\frac{n}{2}}} {\sqrt{\pi}x^{1+\frac{n}{2}}} \Gamma{(\frac{1+n}{2})} 
    \frac{(2 \ell +1+n)(\ell+n)!}{\ell ! n!} e^{i \delta_{n,\ell}}
                                                        \\   \nonumber
& & \times \sin \left[x + i \lambda_n \ln|x-x_+|- \frac{\pi}{2}(\ell+\frac{n}{2})
+ \delta_{n,\ell} \right] + O\left(\frac{1}{x^{2+ \frac{n}{2}}}\right).
\end{eqnarray}
Following the same way of section II the Wronskian $W[R_{n,\ell}^*, R_{n,\ell}]_x$
becomes in the form
\begin{eqnarray}
\label{blwronskian3}
W[R_{n,\ell}^*, R_{n,\ell}]_x &=& -i \frac{2^n}{\pi} \Gamma^2{(\frac{1+n}{2})}
\left( \frac{(2 \ell+1+n)(\ell+n)!}{\ell ! n!} \right)^2
                                                         \\   \nonumber
&\times &\frac{x^n e^{-2 \beta_{n,\ell}}}{(x^{n+1}-x_+^{n+1})(x^{n+1}-x_-^{n+1})}
\sinh{2 \beta_{n,\ell}}
\end{eqnarray}
where we assumed $\delta_{n,\ell}$ is a complex quantity, {\it i.e.} 
$\delta_{n, \ell} \equiv \eta_{n, \ell}+ i \beta_{n, \ell}$. Since $R_{n,\ell}$ and
${\cal G}_{n, \ell}$ exhibit a same behavior around the near-horizon region,
$W[R_{n,\ell}^*, R_{n,\ell}]_x$ should be same with 
$W[{\cal G}_{n, \ell}^*, {\cal G}_{n, \ell}]_x$, which is given in 
Eq.(\ref{blwronskian2}). Equating those two Wronskians yield a relation.
\begin{equation}
\label{blgell}
|g_{n,\ell}|^2 = \frac{2^{n-2}}{\pi x_+^{n+2}} \Gamma^2{(\frac{1+n}{2})}
\left(\frac{(2 \ell + 1+n)(\ell+n)!}{\ell ! n!} \right)^2 (1-e^{-4 \beta_{n,\ell}})
\end{equation}
  
As in the same way with section II we introduce $\tilde{R}_{n,\ell}(x, x_+, x_-)$, 
which is different from $R_{n,\ell}(x, x_+,x_-)$ in its normalization in
such a way that
\begin{equation}
\label{bltilda1}
\tilde{R}_{n,\ell}(x, x_+, x_-)\stackrel{x \rightarrow x_+}{\sim}
(x - x_+)^{\lambda_n}\left[ 1 + O (x - x_+)\right].
\end{equation}
Expressing $\tilde{R}_{n,\ell}$ in terms of the jost functions
\begin{equation}
\label{bltilda2}
\tilde{R}_{n,\ell}(x, x_+, x_-) = f_{n,\ell}^{(-)}(x_+, x_-) 
{\cal F}_{n,\ell(+)} (x, x_+, x_-) +
f_{n, \ell}^{(+)}(x_+, x_-){\cal F}_{n,\ell(-)}(x, x_+, x_-),
\end{equation}
one can show easily that the jost functions $f_{n,\ell}^{(\pm)}$ are obtained by
\begin{eqnarray}
\label{bljost1}
f_{n,\ell}^{(\pm)}(x_+, x_-) &=& \pm \frac{(x^{n+1}-x_+^{n+1})(x^{n+1}-x_-^{n+1})}
{2i x^{n}} W[{\cal F}_{n,\ell (\pm)}, \tilde R]_x
                                                       \\   \nonumber
&=& \pm \frac{\omega^{n+1} (r^{n+1} - r_+^{n+1}) (r^{n+1} - r_-^{n+1})}{2i r^{n}}
     W[{\cal F}_{n,\ell (\pm)}, \tilde R]_r.
\end{eqnarray}
Inserting the explicit from of ${\cal F}_{n,\ell(\pm)}$ into Eq.(\ref{bltilda2})
and comparing it with Eq.(\ref{blrsolution1}), one can derive the relations
\begin{eqnarray}
\label{blscattering1}
S_{n, \ell} (x_+, x_-)&=&\frac {f_{n, \ell}^{(+)}(x_+, x_-)}  
                               {f_{n, \ell}^{(-)}(x_+, x_-)} 
                                              \\  \nonumber
f_{n, \ell}^{(-)}(x_+, x_-)&=&\frac{2^{\frac{n}{2}-1}}{\sqrt{\pi} g_{n,\ell}(x_+,x_-)}
\Gamma{(\frac{1+n}{2})} \frac{(2 \ell+1+n)(\ell+n)!}{\ell ! n!}.
\end{eqnarray}
Combining Eq.(\ref{blgell}) and (\ref{blscattering1}) enables us to express the
greybody factor in terms of the jost function
\begin{equation}
\label{bljost2}
1 - |S_{n,\ell}(x_+, x_-)|^2 = 
\frac{x_+^{n+2}}{|f_{n,\ell}^{(-)} (x_+, x_-)|^2}.
\end{equation}
Thus making use of Eq.(\ref{partialabs}) with replacing ${\cal T}_\ell(\omega)$ 
by $1-|S_{n,\ell}(x_+,x_-)|^2$, the absorption cross section for the bulk scalar is expressed 
in the form
\begin{equation}
\label{blsection2}
\sigma_{n,\ell}^{BL} = 2^{n+1} \pi^{\frac{n+1}{2}} \Gamma{(\frac{n+3}{2})}
\frac{(2 \ell+n+1)(\ell+n)!}{(n+1)! \ell !} \frac{r_+^{n+2}}{|f_{n,\ell}^{(-)}|^2}.
\end{equation}

Therefore we can compute the absorption spectrum completely if the jost function
$f_{n,\ell}^{(-)}$ is computed. Once the absorption cross section is known, 
the emission 
spectrum is also computed with an aid of the Hawking formula (\ref{hawking1}).
 
\section{Absorption and Emission for the bulk scalar}
In this section we will compute the jost functions $f_{n,\ell}^{(\pm)}(x_+, x_-)$ 
introduced in section IV (see Eq.(\ref{bltilda2})) numerically. Following the same way
of section III we firstly derive a solution $\varphi_{n,\ell}(x, x_+,x_-)$ of the 
radial equation (\ref{blradial1}), which has a convergent range in the neighborhood
of an arbitrary point $x=b$. Of course this solution is expressed as a power 
series in the form
\begin{equation}
\label{power}
\varphi_{n, \ell}(x, x_+, x_-) = (x-x_+)^{\lambda_n} \sum_{N=0}^{\infty} D_N (x-b)^N.
\end{equation}
Of course, when $n=0$, the recursion relation of $D_N$ is exactly same with
Eq.(\ref{contin2}) because there is no distinction between brane and bulk.
The recursion relation for $n=1$ is given explicitly in appendix B. Then, as 
explained in section II, it is possible to compute the jost functions 
{\it via} the
analytic continuation.

\begin{figure}[ht!]
\begin{center}
\epsfysize=6.5 cm \epsfbox{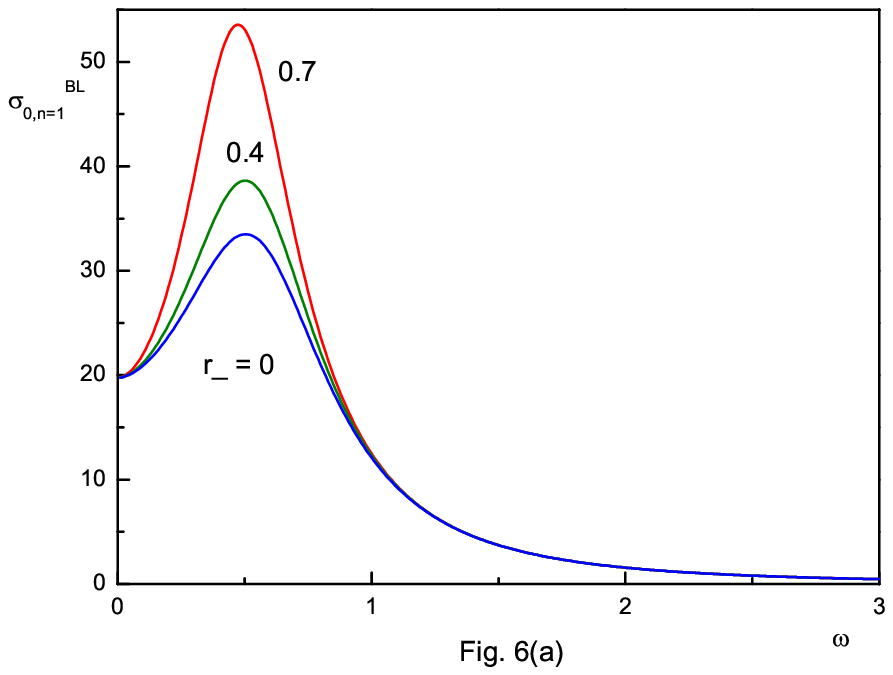}
\epsfysize=6.5 cm \epsfbox{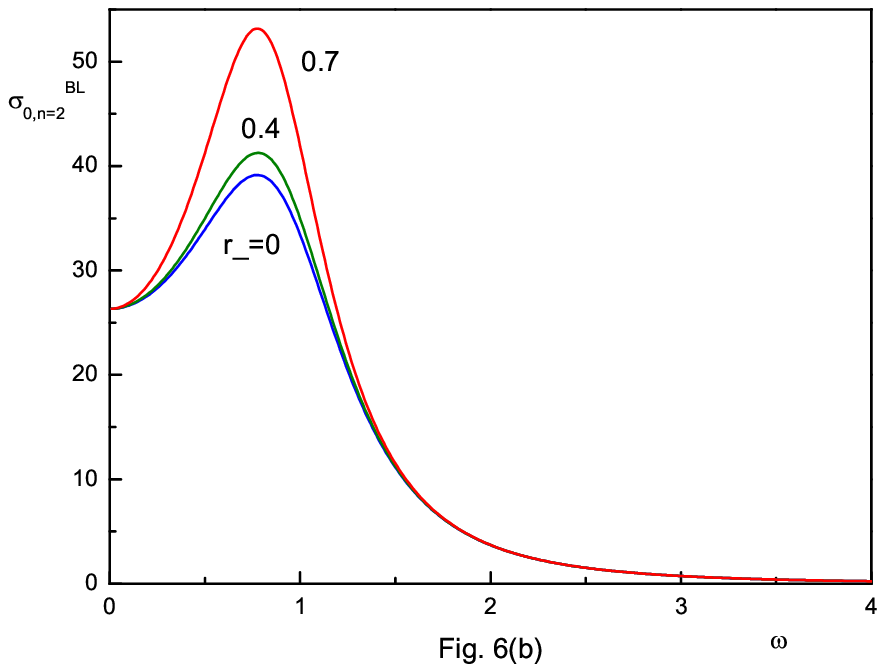}
\caption[fig6]{Plot of the partial absorption cross section for 
s-wave ($\ell = 0$) for 
$n = r_+ = 1$ (Fig. 6(a)) and $n = 2r_+ = 2$ (Fig. 6(b)) with varying $r_-$. Plot
for $n=0$ is same with Fig. 2(a). These figures show how the absorption cross section
is enhanced with increasing $r_-$. Regardless of $r_-$, the low-energy cross sections 
exactly equal to the area of the horizon hypersurface.}
\end{center}
\end{figure}

Fig. 6 shows $r_-$-dependence of the partial absorption cross section for $s$-wave 
($\ell=0$) when $n=1$ (Fig. 6(a)) and $n=2$ (Fig. 6(b)). For the case of $n=0$
Fig. 2(a) should be reproduced as remarked earlier. Like a brane-localized case 
increasing $r_-$ enhances the absorption cross section, which can be understood 
from the Hawking temperature (\ref{mqst}). Unlike a brane-localized case, however, 
the low-energy absorption cross sections are $2 \pi^2 r_+^3$ for $n=1$ and 
$8 \pi^2 r_+^4/3$ for $n=2$, which coincide with the area of the horizon
hypersurfaces. While sharp peak disappears in Fig. 2 for large $n$, this 
property is not maintained in the bulk absorption problem.

\begin{figure}[ht!]
\begin{center}
\epsfysize=6.3 cm \epsfbox{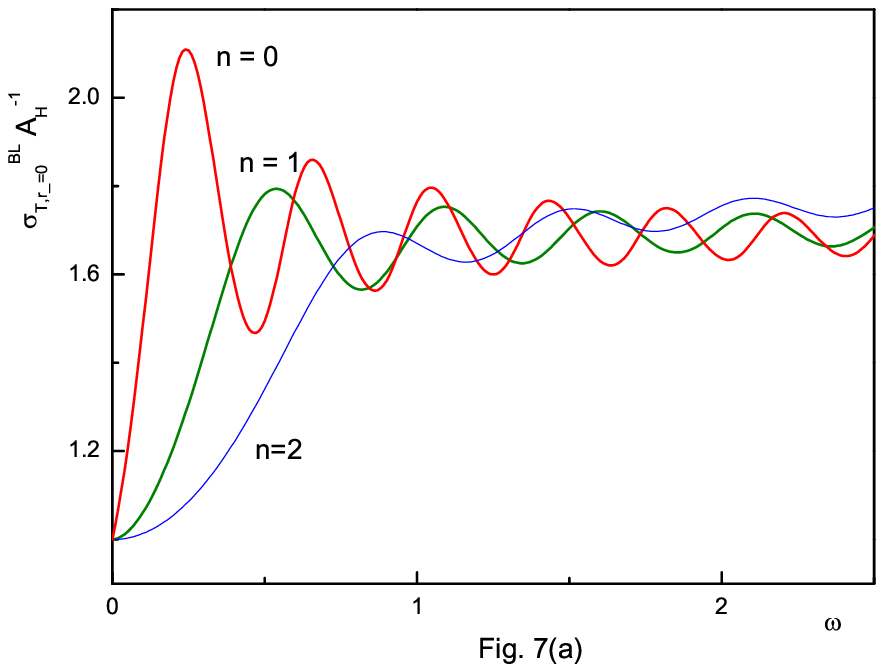}
\epsfysize=6.3 cm \epsfbox{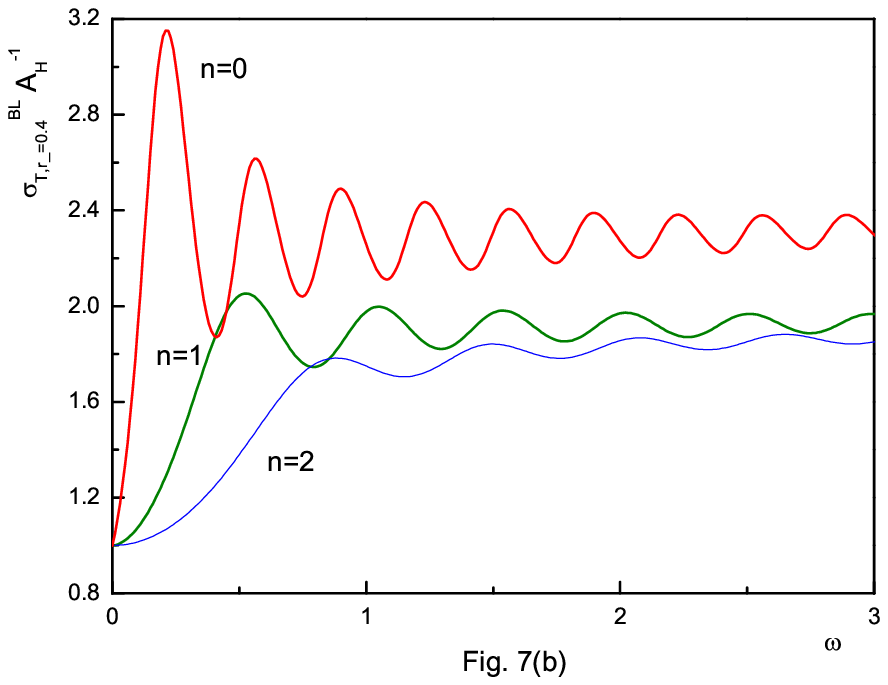}
\epsfysize=6.3 cm \epsfbox{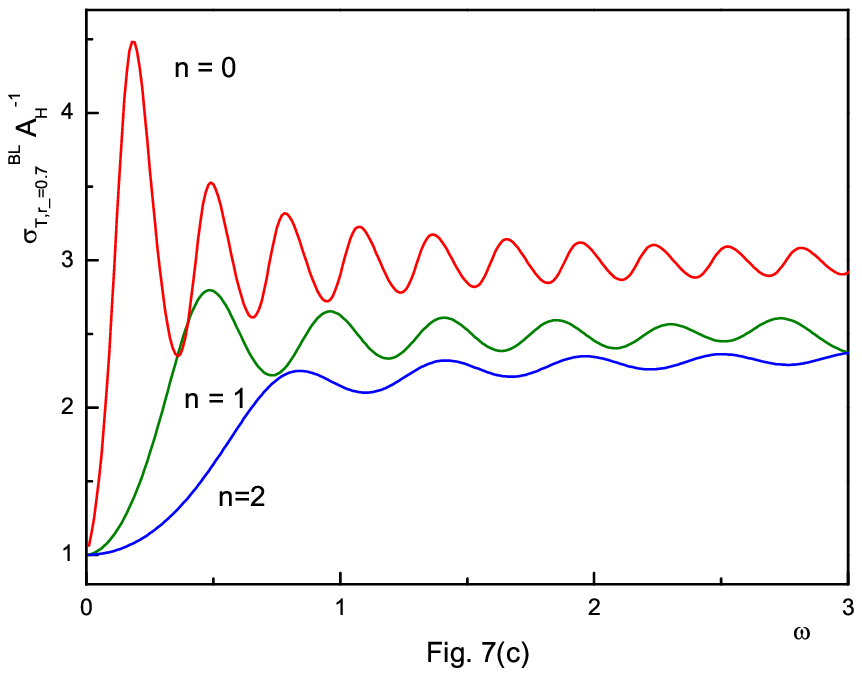}
\caption[fig6]{Plot of $\sigma_T^{BL} / A_H$ when $r_+ = 1$, and $r_- = 0$ 
(Fig. 7(a)), $r_- = 0.4$ (Fig. 7(b)) and $r_- = 0.7$ (Fig. 7(c)) with varying $n$. 
Unlike Fig. 3 the oscillatory pattern does not disappear in the presence of 
nonzero $n$. For nonzero $n$ the total absorption cross section tends to be inclined 
with positive slope. This slope seems to increase with increasing $n$.}
\end{center}
\end{figure}

Fig. 7 shows the $n$-dependence of $\sigma_T^{BL}/A_H$ when $r_-=0$ (Fig. 7(a)), $0.4$
(Fig. 7(b)) and $0.7$ (Fig. 7(c)) where $\sigma_T^{BL}$ and $A_H$ are total absorption
cross section and the horizon area, respectively. Unlike Fig. 3 for the brane-localized 
case the oscillatory behavior does not disappear regardless of $n$ in spite of 
the decrease of the oscillation amplitude. This
fact indicates that the presence of the extra dimensions does not strongly suppress the
absorptivity, which can be understood from Fig. 6. A remarkable fact 
appearing in Fig. 7 
is that the total absorption cross section tends to be inclined with a positive
slope in the course of 
oscillation 
when the extra dimensions exist. The slope seems to increase with increasing $n$. 
Similar behavior was found in the absorption of the dilaton-axion by an extremal 
D3-brane \cite{cve01-1}.

\begin{figure}[ht!]
\begin{center}
\epsfysize=6.0 cm \epsfbox{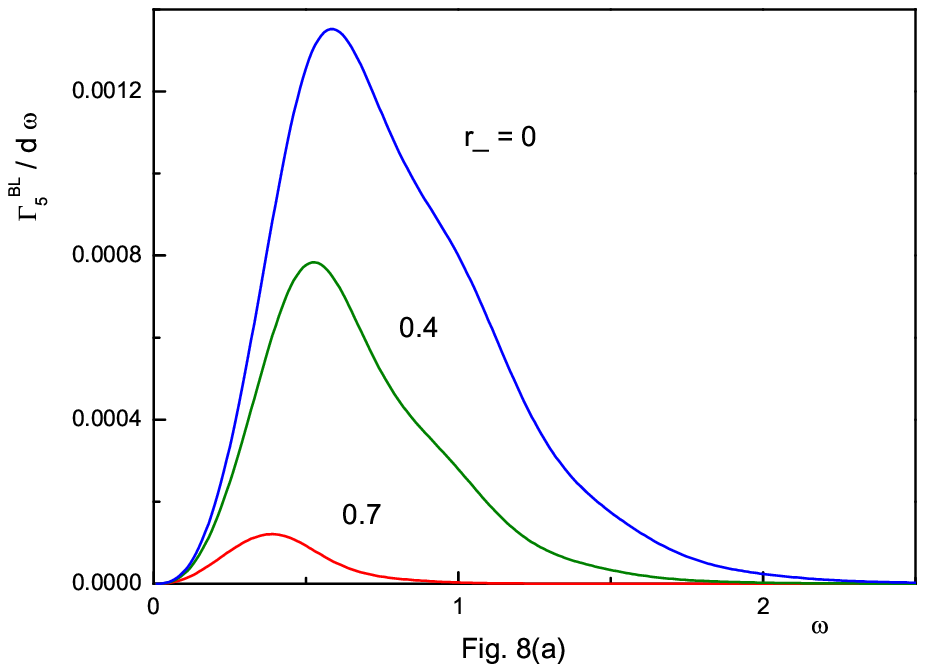}
\epsfysize=6.0 cm \epsfbox{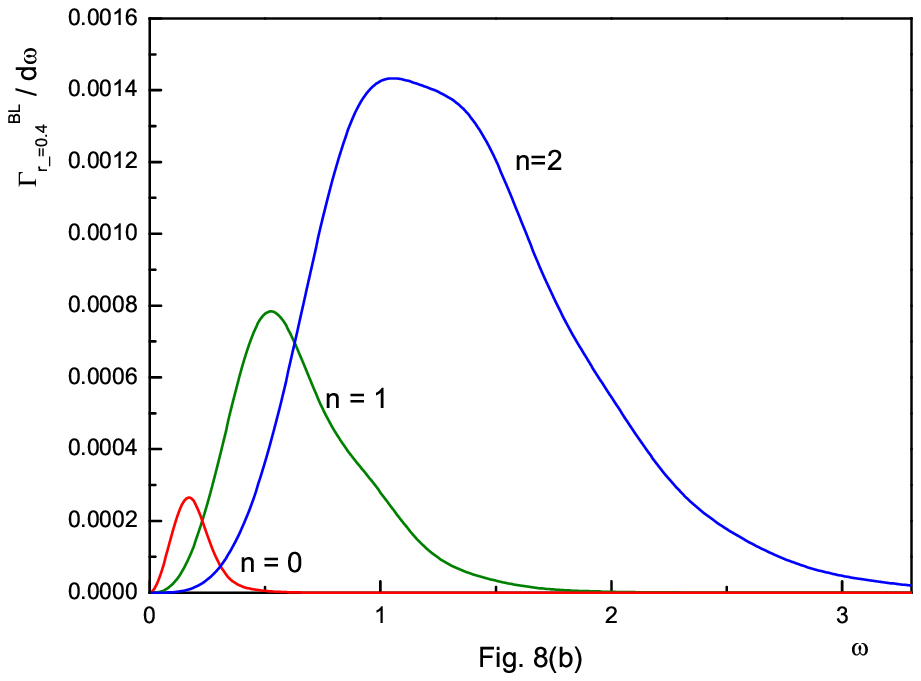}
\caption[fig8]{Plot of the emission rate when $r_+ = n = 1$ with varying $r_-$ 
(Fig. 8(a)), and when $r_+ = 1$ and $r_- = 0.4$ with varying $n$ (Fig. 4(b)). 
Fig. 8(a) indicates that the emission rate is suppressed with increasing $r_-$. 
The presence of the extra dimensions generally enhances the emission rate as 
Fig. 8(b) indicates. These facts imply that the Planck factor is a dominant factor
in the emission spectrum.}
\end{center}
\end{figure}

Finally let us discuss the emission problem for the bulk scalar. The Hawking 
formula (\ref{hawking1}) makes the bulk emission, {\it i.e.} the energy emitted
to the bulk per unit time and energy interval $d\omega$ to be
\begin{equation}
\label{blemi}
\Gamma_D^{BL} = \left( 2^{D-2} \pi^{\frac{D-1}{2}} \Gamma{(\frac{D-1}{2})} \right)^{-1}
\frac{\omega^{D-1} \sigma_{abs}(\omega)}{e^{\frac{\omega}{T_H}}-1} d\omega
\end{equation}
where $D$ is a spacetime dimensions. Since $T_H$ decreases with increasing of $r_-$,
it is evident that the emission rate drastically decreases with increase of $r_-$.
This is confirmed in Fig. 8(a) where $D$ and $r_+$ are fixed as $r_+=1$ and 
$D=5$.
Since, in addition, $T_H$ increases in the presence of the extra dimensions,
the emission rate should be enhanced with increasing $n$. This fact is also confirmed
in Fig. 8(b), where $r_+$ and $r_-$ are fixed as $r_+=1$ and $r_-=0.4$.

\section{Bulk versus Brane}
In this section we would like to compare the emission spectra for the 
brane-localized and the bulk scalars, which was a main issue in 
Ref.\cite{emp00,argy98,banks99}. From Eq.(\ref{hawking2}) and Eq.(\ref{blemi})
the ratio of the emission spectra is 
\begin{equation}
\label{ratio1}
\gamma_D(\omega) \equiv \frac{\Gamma_D^{BL} / d \omega}
                             {\Gamma_D^{BR} / d \omega}
= \frac{\pi^{\frac{5-D}{2}} \omega^{D-4}}{2^{D-3} \Gamma \left( \frac{D-1}{2} \right)}
  \frac{\sigma_T^{BL}}{\sigma_T^{BR}}.
\end{equation}
As expected $\gamma_{4} =1$ when $D=4$ due to $\sigma_T^{BL} = \sigma_T^{BR}$
when $n=0$. When $D=5$ and $6$, this ratio factor becomes
\begin{equation}
\label{ratio2}
\gamma_5(\omega) = \frac{\omega}{4} \frac{\sigma_T^{BL}}{\sigma_T^{BR}}
\hspace{2.0cm}
\gamma_6(\omega) = \frac{\omega^2}{6 \pi} \frac{\sigma_T^{BL}}{\sigma_T^{BR}}.
\end{equation}
Thus in the low-energy region ($\omega << 1$) 
$\gamma_5 \sim \pi r_+ \omega / 8 \sim 0$ and
$\gamma_6 \sim r_+^2 \omega^2 / 9 \sim 0$. Thus in this region the emission into the 
brane is dominant compared to the emission into the bulk. 

In the high-energy region ($\omega >> 1$), however, the emission into the bulk becomes
larger than that into the brane due to $\omega$ factor in Eq.(\ref{ratio1}). 

\begin{figure}[ht!]
\begin{center}
\epsfysize=6.5 cm \epsfbox{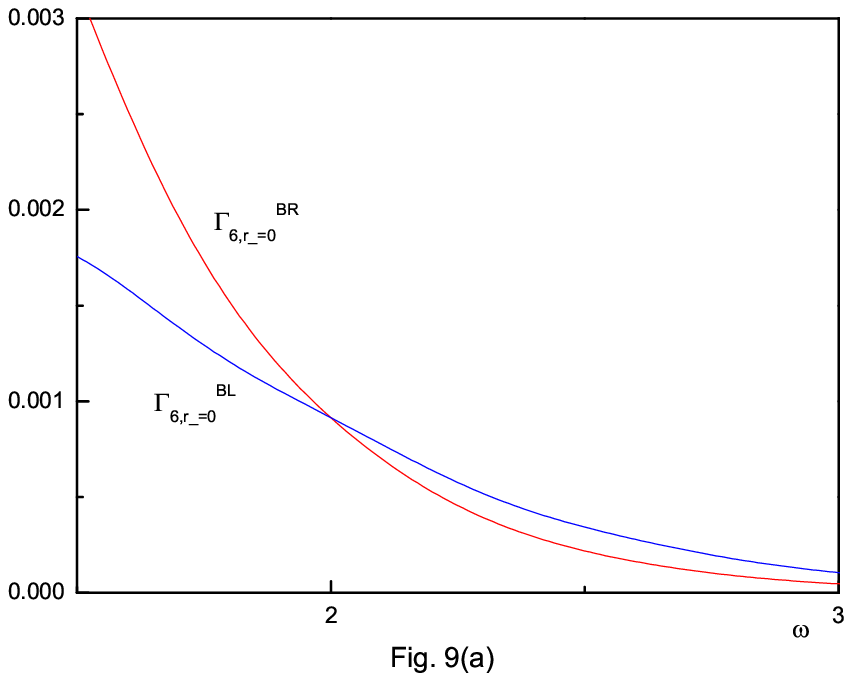}
\epsfysize=6.5 cm \epsfbox{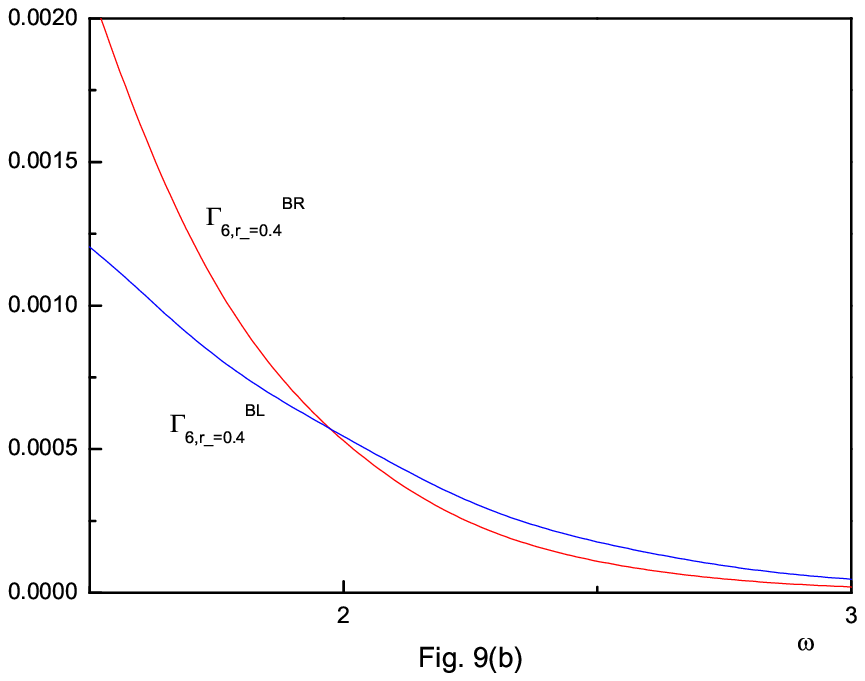}
\epsfysize=6.5 cm \epsfbox{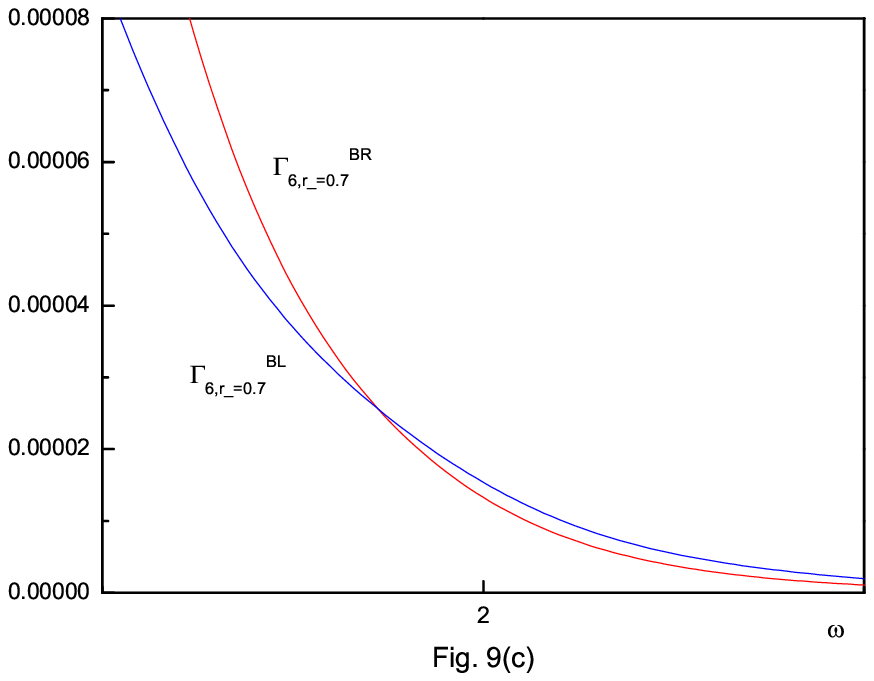}
\caption[fig9]{Plot of the emission rate for the bulk and brane-localized scalars at
$D=6$ in the neighborhood of $\omega_{\ast}$, where 
$\Gamma_6^{BR}(\omega_{\ast}) = \Gamma_6^{BL}(\omega_{\ast})$, when 
$r_- = 0$ (Fig. 9(a)), $r_- = 0.4$ (Fig. 9(b)) and $r_- = 0.7$ (Fig. 9(c)). These
figures show that $\Gamma_6^{BR} > \Gamma_6^{BL}$ when $\omega < \omega_{\ast}$
and $\Gamma_6^{BR} < \Gamma_6^{BL}$ when $\omega > \omega_{\ast}$. This fact indicates
that the emission into the brane is dominant in the domain of low-energy and the 
emission into bulk is dominant in the domain of high-energy.}
\end{center}
\end{figure}

The low-energy and high-energy behavior of $\gamma_D(\omega)$ are confirmed in
Fig. 9, where the emission spectra for the bulk and the brane-localized scalars
are plotted simultaneously for $r_- = 0$ (Fig. 9(a)), $r_- = 0.4$ (Fig. 9(b))
and $r_- = 0.7$ (Fig. 9(c)) when $D=6$. 
These figures are plotted in the neighborhood of a point where $\Gamma_6^{BR}$ and 
$\Gamma_6^{BL}$ coincide with each other.
These figures shows that $\Gamma_6^{BL} - \Gamma_6^{BR}$ in the
high-energy domain decreases with increasing $r_-$.  
This means the ratio of the total 
emission rate $\Gamma_D^{BL} / \Gamma_D^{BR}$ decreases with increasing $r_-$.

To confirm this fact we computed $\Gamma_D^{BL}$, $\Gamma_D^{BR}$ and their
ratio for $r_- = 0$, $r_- = 0.4$, and $r_- = 0.7$ when $n = 0$ (Table 1), 
$n = 1$ (Table 2), and $n = 2$ (Table 3).

\begin{center}
{\large{Table1}}: Comparision of bulk to brane emission in $n=0$
\end{center}

\begin{center}
%\begin{tabular}{|c|c|c|c|}
\begin{tabular}{cccc}
\hline
$     $                &\hspace{0.5cm} $r_-=0$    &\hspace{0.5cm} $r_-=0.4$   
                                                  &\hspace{0.5cm} $r_-=0.7$ \\
\hline\hline
$\Gamma_4^{BR}$      &\hspace{0.5cm} $0.000297531$   &\hspace{0.5cm} $0.0000522908$
                                 & \hspace{0.5cm} $3.30229 \times 10^{-6}$ \\
\hline
$\Gamma_4^{BL}$      &\hspace{0.5cm} $0.000297531$   &\hspace{0.5cm} $0.0000522908$
                                        &\hspace{0.5cm} $3.30229 \times 10^{-6}$ \\
\hline
$\Gamma_4^{BL}/\Gamma_4^{BR}$   &\hspace{0.5cm} $1$          &\hspace{0.5cm} $1$
                                                       &\hspace{0.5cm} $1$ \\
\hline\hline
\end{tabular}
\end{center}

\vspace{1.0cm}
\begin{center}
{\large{Table2}}: Comparision of bulk to brane emission in $n=1$
\end{center}

\begin{center}
\begin{tabular}{cccc}
\hline
$     $                &\hspace{0.5cm} $r_-=0$        &\hspace{0.5cm} $r_-=0.4$
                                               &\hspace{0.5cm} $r_-=0.7$ \\
\hline\hline
$\Gamma_5^{BR}$      &\hspace{0.5cm} $0.00266319$   &\hspace{0.5cm} $0.00141917$ 
                                     &\hspace{0.5cm} $0.000223692$ \\
\hline
$\Gamma_5^{BL}$      &\hspace{0.5cm} $0.00104967$   &\hspace{0.5cm} $0.000488455$
                                    &\hspace{0.5cm} $0.0000478151$ \\
\hline
$\Gamma_5^{BL}/\Gamma_5^{BR}$   &\hspace{0.5cm} $0.394$    &\hspace{0.5cm} $0.344183$
                                        &\hspace{0.5cm} $0.213754$ \\
\hline\hline
\end{tabular}
\end{center}

\vspace{1.0cm}
\begin{center}
{\large{Table3}}: Comparision of bulk to brane emission in $n=2$
\end{center}

\begin{center}
\begin{tabular}{cccc}
\hline
$     $                &\hspace{0.5cm} $r_-=0$        &\hspace{0.5cm} $r_-=0.4$  
                                               &\hspace{0.5cm} $r_-=0.7$ \\
\hline\hline
$\Gamma_6^{BR}$      &\hspace{0.5cm} $0.0107479$   &\hspace{0.5cm} $0.00841889$  
                           &\hspace{0.5cm} $0.00224066$ \\
\hline
$\Gamma_6^{BL}$      &\hspace{0.5cm} $0.00267917$   &\hspace{0.5cm} $0.00187831$  
                                     &\hspace{0.5cm} $0.000263698$ \\
\hline
$\Gamma_6^{BL}/\Gamma_6^{BR}$   &\hspace{0.5cm} $0.249274$    &\hspace{0.5cm} 
                        $0.223107$    &\hspace{0.5cm} $0.117688$ \\
\hline\hline
\end{tabular}
\end{center}

These Tables indicate that the emission into the brane is dominant compared to 
emission into the bulk regardless of $r_-$ in the presence of extra dimensions,
which is a main result of Ref.\cite{emp00}.
The ratio factor $\Gamma_D^{BL} / \Gamma_D^{BR}$ decreases with increasing $r_-$
as expected. Although the calculation is not explicitly carried out in this paper,
the emission into bulk might be dominant compared to the emission into brane if our
universe has sufficiently many extra dimensions. This is conjectured from the 
fact that the 
dominance of bulk emission in the high-energy domain can overwhelm the dominance
of brane emission in the low-energy domain if $n$ is a sufficiently large due to 
the factor $\omega^{D-4}$ in Eq.(\ref{ratio1}).

Although our numerical computation supports the main claim of Ref.\cite{emp00}, 
{\it i.e. black holes radiate mainly on the brane}, there are several issues to 
consider seriously
before we arrive at a conclusion for the dominant emissivity.
Firstly, one should note that our computation presented in the Tables is restricted
to the relative emission for the spin-$0$ field. Since the emission is in general
highly dependent on the field spin, our result does not derive any conclusion to
the dominant one between emission on the brane and off the brane when fields
have higher spin. The emission for the higher spin particles manifestly 
need another detailed calculation.

Secondly, we have not considered in this paper the rotating effect of the 
black holes. Since the black holes produced by high-energy scattering are 
expected to be highly rotating due to non-zero impact parameter, it is
very important to 
consider this effect. It is well-known that there exist the superradiance
modes\cite{zeldo71-1,press72} in the general relativistic rotating 
black holes, which means the amplification of the amplitude of the 
incident wave by making use of the energy extracted from a black hole.
Recently, the existence of the superradiance modes in the brane world
black holes has been shown analytically\cite{frol03-1} and 
numerically\cite{ida05,harris05-1}. Also importance of the superradiance
for the experimental signature
in the colliders is stressed in Ref.\cite{frol02-1,frol02-2}. In 
Ref.\cite{frol03-1} the existence of the superradiance for the bulk
scalars incident on five-dimensional rotating black holes is proved
analytically. In Ref.\cite{ida05,harris05-1} the effect of the 
superradiance for the brane-localized scalars has been studied numerically
at $D=10$ and $D=6$, respectively. Since the effect of the superradiance 
is not taken into account in Ref.\cite{emp00} and present paper, the debate
on the dominant emissivity is not finished yet.  

\section{Conclusion}
In this paper we examined the absorption and emission spectra for the brane-localized
and bulk scalars when the spacetime is a ($4+n$)-dimensional 
RN black hole. In particular, the effects of the inner 
horizon parameter $r_-$ and the number of toroidally compactified extra dimensions $n$
are rigorously discussed throughout the paper. For calculational technique we adopt
the numerical calculation introduced in Ref\cite{sanc78,sanc98,jung04-2} for the 
computation of the absorption and emission spectra in the full-range of energy
of the scalar particles. 

For the case of the brane-localized scalar it turns out that the presence of 
$r_-$ generally enhances the absorptivity compared to the case of the Schwarzschild 
phase. This can be deduced from a fact that the Hawking temperature given in
Eq.(\ref{mqst}) decreases with increasing $r_-$. The low-energy absorption cross 
section exactly equals to $4 \pi r_+^2$, which is an universality for the 
spherically symmetric and asymptotically flat black holes. The emission is generally
reduced with increasing $r_-$, which indicates the the Planck factor is more
dominant than the greybody factor in the emission problem. The presence of the 
extra dimensions suppresses the absorption spectrum and enhances the emission 
spectrum compared to those without the extra dimension. Especially, the oscillatory 
pattern of the total absorption cross section disappears when $n$ is large.  
The reason for this might be the fact that the suppression of the absorptivity
in the presence of the extra dimensions is too strong. 

For the case of the bulk scalar the effects of $r_-$ and $n$ 
in the absorption and emission spectra are similar to 
the case of the brane-localized scalar. Unlike the brane-localized case, however,
the disappearance of the oscillatory pattern in the total absorption cross section
does not manifestly happen in this case although the amplitude of oscillation 
decreases with increasing $n$. 
This means the presence of the extra dimensions 
does not suppress the absorptivity too much strongly. 
The total absorption cross section tends to be inclined with a positive slope when
the extra dimensions exist. This slope seems to increase with increasing $n$.
The low-energy absorption cross section is 
not dependent on $r_-$ and equals to the area of the horizon hypersurface, which
is also universality for the spherically symmetric and asymptotically flat black
holes. 

Finally, we discussed the ratio of the brane emission to the bulk emission. It turns
out that the brane emission is dominant in the low-energy domain while the bulk
emission is larger than the brane emission in the high-energy domain. For 
comparatively small $n$ the total emission rate into the brane is much larger 
than that into the bulk. If, however, $n$ is sufficiently large, the bulk emission
can be dominant because of $\omega$-dependent factor in Eq.(\ref{ratio1}). Our 
numerical calculation shows that the ratio factor $\Gamma_D^{BL} / \Gamma_D^{BR}$ 
decreases with increasing $r_-$.

It is interesting to extend our paper to the particles with spin. This is important
when $n$ is large because the emission rate for the particles with higher spin 
is dominant in the higher-dimensional theories. We may use a 
Newman-Penrose formalism\cite{newman62,chand83} to derive a master equation. It is 
interesting to discuss the effect of the inner horizon parameter $r_-$ in  
the absorption and emission problems for the particles with spin.

Although this paper is strongly motivated by the recent brane-world scenarios,
our computational techniques can be directly applied to the five-dimensional 
RN black hole carrying three different electric charges and the corresponding 
six-dimensional black string introduced in Ref.\cite{mal96-1,hawk97} for the 
computation of the absorption and emission spectra in the full-range of energy.
This calculation may give interesting results in the correspondence between the 
black hole and D-brane beyond the dilute gas region.

It is also important to extend our calculation to the rotating effect of 
black holes. Especially, as claimed in Ref.\cite{frol03-1,frol02-1,frol02-2}
the effect of the 
superradiance should be examined carefully 
to determine the dominant one between bulk and brane emission. Work on this
issue is in progress and will be reported elsewhere.

\vspace{1cm}

{\bf Acknowledgement}:  
This work was supported by the Kyungnam University
Research Fund, 2004.

\newpage
\begin{appendix}{\centerline{\bf Appendix A}}
\setcounter{equation}{0}
\renewcommand{\theequation}{A.\arabic{equation}}
The recursion relation for $n=1$ which is correspondent to 
Eq.(\ref{recursion1}) for $n=0$ is 
\begin{eqnarray}
\label{appena1}
& &\left[(N + \lambda_1) (N + \lambda_1 - 1) \alpha_1 + (N + \lambda_1) \beta_1
      + \gamma_1 \right] d_{\ell,N} 
                                                   \\   \nonumber
&+ & \left[(N + \lambda_1 - 1) (N + \lambda_1 - 2) \alpha_2 + (N + \lambda_1 - 1)
    \beta_2
      + \gamma_2 \right] d_{\ell,N-1} 
                                                   \\   \nonumber
&+ & \left[(N + \lambda_1 - 2) (N + \lambda_1 - 3) \alpha_3 + (N + \lambda_1 - 2)
        \beta_3 + \gamma_3 \right] d_{\ell,N-2} 
                                                   \\   \nonumber
&+ &  \left[(N + \lambda_1 - 3) (N + \lambda_1 - 4) \alpha_4 + (N + \lambda_1 - 3)
                               \beta_4
      + \gamma_4 \right] d_{\ell,N-3} 
                                                   \\   \nonumber
&+ & \left[(N + \lambda_1 - 4) (N + \lambda_1 - 5) \alpha_5 + (N + \lambda_1 - 4)
                               \beta_5
      + \gamma_5 \right] d_{\ell,N-4} 
                                                   \\   \nonumber
&+ & \left[(N + \lambda_1 - 5) (N + \lambda_1 - 6) \alpha_6 + (N + \lambda_1 - 5)
                               \beta_6
      + \gamma_6 \right] d_{\ell,N-5} 
                                                   \\   \nonumber
&+ & \left[(N + \lambda_1 - 6) (N + \lambda_1 - 7) \alpha_7 + (N + \lambda_1 - 6)
                               \beta_7
      + \gamma_7 \right] d_{\ell,N-6} 
                                                   \\   \nonumber
&+ & \left[(N + \lambda_1 - 7) (N + \lambda_1 - 8) + 2 (N + \lambda_1 - 7)
      + \gamma_8 \right] d_{\ell,N-7} + \gamma_9 d_{\ell, N-8} + 
       d_{\ell, N-9} = 0
\end{eqnarray}
where             
\begin{eqnarray}
\label{appena2}
& &\alpha_1 = 4 x_+^3 (x_+^2 - x_-^2)^2 \hspace{3.2cm}
   \alpha_2 = 8 x_+^2 (x_+^2 - x_-^2) (3 x_+^2 - x_-^2) 
                                                       \\  \nonumber
& &\alpha_3 = x_+ (61 x_+^4 - 50 x_+^2 x_-^2 + 5 x_-^4) \hspace{1.0cm}
\alpha_4 = 85 x_+^4 - 38 x_+^2 x_-^2 + x_-^4  
                                                       \\  \nonumber
& &\alpha_5 = 14 x_+ (5 x_+^2 - x_-^2)    \hspace{2.30cm}
   \alpha_6 = 2 (17x_+^2 - x_-^2)         \hspace{1.0cm}
   \alpha_7 = 9 x_+  
                                                       \\  \nonumber
& &\beta_1 = 4 x_+^3 (x_+^2 - x_-^2)^2
   \hspace{2.5cm}   \beta_2 = 2 x_+^2 (x_+^2 - x_-^2) (13 x_+^2 - x_-^2)
                                                       \\   \nonumber
& &\beta_3 = 8 x_+^3 (9 x_+^2 - 5 x_-^2)   \hspace{2.3cm}
   \beta_4 = 10 x_+^2 (11 x_+^2 - 3 x_-^2) 
                                                       \\    \nonumber
& & 
   \beta_5 = 4 x_+ (25 x_+^2 - 3 x_-^2) \hspace{2.3cm}
   \beta_6 = 2 (27 x_+^2 - x_-^2)  \hspace{1.0cm}
   \beta_7 = 16 x_+    
                                                        \\    \nonumber
& &
   \gamma_1 = x_+^9       \hspace{5.0cm}
   \gamma_2 = x_+^4 \left[9 x_+^4 - 2 \ell (\ell + 1) (x_+^2 - x_-^2) \right]
\hspace{1.0cm}
                                                        \\    \nonumber
& &
   \gamma_3 = x_+^3 \left[36 x_+^4 - \ell (\ell + 1) (11 x_+^2 - 7 x_-^2) \right]
   \hspace{1.0cm}
   \gamma_4 = x_+^2 \left[84 x_+^4 - \ell (\ell + 1) (25 x_+^2 - 9 x_-^2) \right]
                                                         \\    \nonumber
& &
   \gamma_5 = x_+ \left[126 x_+^4 - 5 \ell (\ell + 1) (6 x_+^2 - x_-^2) \right]
\hspace{1.0cm}
   \gamma_6 = 126 x_+^4 - \ell (\ell + 1) (20 x_+^2 - x_-^2)
                                                         \\    \nonumber
& &
   \gamma_7 = 7 x_+ \left[12 x_+^2 - \ell (\ell + 1)\right]
\hspace{1.0cm}
   \gamma_8 = 36 x_+^2 - \ell (\ell + 1)    
\hspace{1.0cm}
    \gamma_9 = 9 x_+.
\end{eqnarray}

The recursion relation for $n=1$ which is correspondent to Eq.(\ref{recursion2}) 
is
\begin{eqnarray}
\label{appena3}
& & \tau_{N (-)} [-(N+1) \tilde{\beta}_1 + \tilde{\gamma}_1]
  + \tau_{N-1 (-)} [N(N+1) - N \tilde{\beta}_2 + \tilde{\gamma}_2]
                                                         \\  \nonumber
& & + \tau_{N-2 (-)} [-(N-1) \tilde{\beta}_3 + \tilde{\gamma}_3]
    + \tau_{N-3 (-)} [(N-2)(N-1) \tilde{\alpha}_1 - (N-2) \tilde {\beta}_4 
                                + \tilde{\gamma}_4]
                                                         \\  \nonumber
& & + \tau_{N-4 (-)} [-(N-3) \tilde{\beta}_5 + \tilde{\gamma}_5] 
    + \tau_{N-5 (-)} [(N-4)(N-3) \tilde{\alpha}_2 - (N-4) \tilde{\beta}_6 
                                + \tilde{\gamma}_6]
                                                        \\  \nonumber
& & + \tau_{N-6 (-)} [-(N-5) \tilde{\beta}_7 + \tilde{\gamma}_7]
    + \tau_{N-7 (-)} [(N-6)(N-5) \tilde{\alpha}_3 - (N-6) \tilde{\beta}_8
                               + \tilde{\gamma}_8]  
                                                        \\ \nonumber
& & + \tau_{N-8 (-)} [-(N-7) \tilde{\beta}_9 + \tilde{\gamma}_9]
    + \tau_{N-9 (-)} [(N-8)(N-7) \tilde{\alpha}_4 - (N-8) \tilde{\beta}_{10}] = 0
\end{eqnarray}
where $\tau_{0 (-)} = 1$, $\tau_{N (+)} = \tau_{N (-)}^*$ and
\begin{eqnarray}
\label{appena4}
& & \tilde{\alpha}_1 = -2(x_+^2 + x_-^2)    \hspace{5.0cm}
    \tilde{\alpha}_2 = x_+^4 + 4 x_+^2 x_-^2 + x_-^4   
                                                 \\  \nonumber
& &
    \tilde{\alpha}_3 = -2 x_+^2 x_-^2 (x_+^2 + x_-^2)  \hspace{4.0cm}
    \tilde{\alpha}_4 = x_+^4 x_-^4
                                                  \\   \nonumber
& & \tilde{\beta}_1 = 2i    \hspace{2.0cm}
    \tilde{\beta}_2 = 2(1 - \lambda_1)    \hspace{2.0cm}
    \tilde{\beta}_3 = -2[\lambda_1 x_+ + 2i(x_+^2 + x_-^2)]  
                                                  \\    \nonumber
& & \tilde{\beta}_4 = -2[(x_+^2 + x_-^2) - \lambda_1(x_+^2 + 2 x_-^2)] 
    \hspace{1.0cm}                         
    \tilde{\beta}_5 = 2i[(x_+^4 + x_-^4) + 4 x_+^2 x_-^2] 
                             + 2 \lambda_1 x_+ (x_+^2 + 2 x_-^2)
                                                  \\    \nonumber
& &
 \tilde{\beta}_6 = - 2 \lambda_1 x_-^2 (2 x_+^2 + x_-^2)  \hspace{2.5cm}
 \tilde{\beta}_7 = -2 x_+ x_-^2[2i x_+ (x_+^2 + x_-^2)
                     + \lambda_1 (2 x_+^2 + x_-^2)]
                                                  \\    \nonumber
& & 
 \tilde{\beta}_8 = 2 x_+^2 x_-^2 [x_+^2 + (1 + \lambda_1) x_-^2]   \hspace{1.5cm}
 \tilde{\beta}_9 = 2 x_+^3 x_-^4 [i x_+ + \lambda_1]        \hspace{1.5cm}
 \tilde{\beta}_{10} = -2 x_+^4 x_-^4
                                                  \\   \nonumber
& & \tilde{\gamma}_1 = 2i(1 - \lambda_1)   \hspace{3.5cm}
    \tilde{\gamma}_2 = 2(x_+^2 + x_-^2) - \ell(\ell+1) -\lambda_1(1-\lambda_1) 
                                        - 2i \lambda_1 x_+
                                                  \\   \nonumber
& & \tilde{\gamma}_3 = -2i[(x_+^2 + x_-^2) - \lambda_1 (x_+^2 + 2 x_-^2)
                                  + i \lambda_1^2 x_+]
                                                   \\   \nonumber
& & \tilde{\gamma}_4 = -(x_+^4 + x_-^4) - 4 x_+^2 x_-^2 
                       + \ell (\ell +1)(x_+^2 + x_-^2) 
                       + \lambda_1^2 (x_+^2 - 2 x_-^2)
                       + \lambda_1 x_+[2i (x_+^2 + 2 x_-^2) + x_+]
                                                    \\   \nonumber
& & \tilde{\gamma}_5 = - 2 \lambda_1 x_-^2 [i(2 x_+^2 + x_-^2) + 2 \lambda_1 x_+
                                             + x_+]
                                                    \\   \nonumber
& & \tilde{\gamma}_6 = \left\{ x_+^2 [2(x_+^2 + x_-^2) - \ell(\ell + 1)]
                          - \lambda_1^2 (2 x_+^2 - x_-^2) 
                      - \lambda_1[2i x_+(2 x_+^2 + x_-^2) - x_-^2] \right \} x_-^2
                                                    \\   \nonumber
& & \tilde{\gamma}_7 = 2 x_+ x_-^2 (\lambda_1 + i x_+) [(x_+^2 + x_-^2) 
                                            + \lambda_1 x_-^2]  \hspace{1.0cm}
 \tilde{\gamma}_8 = - x_+^2 x_-^4 [x_+^2 - \lambda_1^2 + \lambda_1 (1 - 2i x_+)]
                                                       \\   \nonumber
& &    
    \tilde{\gamma}_9 = - 2 x_+^3 x_-^4 (\lambda_1 + i x_+)
\end{eqnarray}
    
The recursion relation for $n=1$ corresponding to Eq.(\ref{contin2}) for $n=0$ is 
\begin{eqnarray}
\label{appena5}
& & D_N[ N(N-1) \bar{\alpha}_9]+D_{N-1}[(N-1)(N-2) \bar{\alpha}_8+(N-1)\bar{\beta}_9]
                                                \\   \nonumber
&+& D_{N-2}[(N-2)(N-3) \bar{\alpha}_7 + (N-2) \bar{\beta}_8 + \bar{\gamma}_9]
+ D_{N-3}[(N-3)(N-4) \bar{\alpha}_6 + (N-3) \bar{\beta}_7 + \bar{\gamma}_8]
                                                 \\   \nonumber
&+& D_{N-4}[(N-4)(N-5) \bar{\alpha}_5 + (N-4) \bar{\beta}_6 + \bar{\gamma}_7]
+ D_{N-5}[(N-5)(N-6) \bar{\alpha}_4 + (N-5) \bar{\beta}_5 + \bar{\gamma}_6]
                                                 \\   \nonumber
&+& D_{N-6}[(N-6)(N-7) \bar{\alpha}_3 + (N-6) \bar{\beta}_4 + \bar{\gamma}_5]
+ D_{N-7}[(N-7)(N-8) \bar{\alpha}_2 + (N-7) \bar{\beta}_3 + \bar{\gamma}_4]
                                                 \\   \nonumber
&+& D_{N-8}[(N-8)(N-9) \bar{\alpha}_1 + (N-8) \bar{\beta}_2 + \bar{\gamma}_3]
+ D_{N-9}[(N-9)(N-10)  + (N-9) \bar{\beta}_1 + \bar{\gamma}_2]
                                                 \\   \nonumber
&+& D_{N-10} \bar{\gamma}_1 + D_{N-11} = 0
\end{eqnarray}
where
\begin{eqnarray}
\label{appena6}
\bar{\alpha}_1 &=& 9b \hspace{1.0cm} \bar{\alpha}_2 = 36 b^2 -2(x_+^2 + x_-^2)
\hspace{1.0cm} \bar{\alpha}_3 = 14 b[6b^2 -(x_+^2+x_-^2)]
                                                \\   \nonumber
 \bar{\alpha}_4 &=& 126 b^4 + (x_+^4+x_-^4) + 4 x_+^2 x_-^2 -42 b^2(x_+^2+x_-^2)
                                                \\   \nonumber
 \bar{\alpha}_5 &=& b[126 b^4 - 70 b^2 (x_+^2 + x_-^2)
                          + 5(x_+^4 + x_-^4 + 4x_+^2 x_-^2)]
                                                 \\   \nonumber
\bar{\alpha}_6 &=& 2 [42 b^6 - 35 b^4(x_+^2+x_-^2)-x_+^2 x_-^2(x_+^2+x_-^2)
                   + 5 b^2(x_+^4 + x_-^4 + 4 x_+^2 x_-^2)]
                                                 \\   \nonumber
\bar{\alpha}_7 &=& 2 b[18 b^6 - 21 b^4(x_+^2+x_-^2)-3x_+^2 x_-^2(x_+^2+x_-^2)
                     +5 b^2(x_+^4 + x_-^4 + 4 x_+^2 x_-^2)]
                                                 \\   \nonumber
\bar{\alpha}_8 &=& [b^4 - b^2(x_+^2+x_-^2)+x_+^2 x_-^2] [9 b^4 +x_+^2 x_-^2
                                -5 b^2(x_+^2+x_-^2)]
                                                 \\   \nonumber
\bar{\alpha}_9 &=& b[b^4 - b^2(x_+^2+x_-^2)+x_+^2 x_-^2]^2
                                                 \\   \nonumber
\bar{\beta}_1 &=& 2 + 2\lambda_1       \hspace{1.0cm}
\bar{\beta}_2 = 16 b + 2 \lambda_1 (8 b + x_+)
                                                 \\   \nonumber
\bar{\beta}_3 &=& 56 b^2-2(x_+^2+x_-^2)+2 \lambda_1[28 b^2+7 b x_+ -(x_+^2+2x_-^2)]
                                                 \\   \nonumber
\bar{\beta}_4 &=& 2[56 b^3 - 6 b (x_+^2+x_-^2)
                      + \lambda_1[56 b^3 + x_+(21 b^2 -x_+^2-2x_-^2) 
                                                 -6 b(x_+^2 + 2 x_-^2)]]
                                                  \\   \nonumber 
\bar{\beta}_5 &=& 2 \bigg[70 b^4 -15b^2(x_+^2+x_-^2)
                                                  \\   \nonumber
& & \hspace{0.5cm} + \lambda_1[70 b^4 + x_+[35 b^3 - 5 b(x_+^2 + 2 x_-^2)]
                           - 15 b^2(x_+^2 + 2 x_-^2)+ x_-^2(2 x_+^2 +x_-^2)] \bigg]
                                                 \\   \nonumber
\bar{\beta}_6 &=& 2 \bigg[56 b^5 -20 b^3(x_+^2+x_-^2)
                                                 \\   \nonumber
& & \hspace{0.5cm} + \lambda_1[56 b^5 + 35 b^4 x_+ -10 b^2(2b+x_+)(x_+^2 + 2 x_-^2)
                           + x_-^2(4 b + x_+)(2 x_+^2 +x_-^2)] \bigg]
                                                  \\   \nonumber
\bar{\beta}_7 &=& 2\bigg[ 28 b^6 -(15 b^4 - x_+^2 x_-^2)(x_+^2+x_-^2)
                                                    \\   \nonumber
               & & \hspace{0.5cm}+ \lambda_1[28 b^6 +21 b^5 x_+ 
                   - 5 b^3(x_+^2 + 2 x_-^2) (2 x_+ + 3b) -x_+^2 x_-^4 + 3 b x_-^2 
                               (2b + x_+)(2x_+^2 + x_-^2)] \bigg]
                                                     \\   \nonumber
\bar{\beta}_8 &=& 2\bigg[8 b^7 + 2 b x_+^2 x_-^2(x_+^2+x_-^2)-6 b^5 x_-^2 + x_+^2
                                                     \\  \nonumber
              & & \hspace{0.5cm}+\lambda_1[8b^7 +7 b^6 x_+ - 12 b^5 x_-^2
                                           - 5 b^4 x_+ (x_+^2 + 2 x_-^2)
                                           + 2 b x_-^2[b(2x_+^2 + x_-^2)-x_+^2 x_-^2]
                                                   - x_+^3 x_-^4 + x_+^2] \bigg]
                                                     \\   \nonumber
\bar{\beta}_9 &=& 2(b^2 - x_-^2)(b^2- x_+^2)[b^4 - x_+^2 x_-^2 
                   + \lambda_1[b^4 + b^3 x_+ - b^2 x_-^2 -b x_+ x_-^2]]
                                                     \\   \nonumber
\bar{\gamma}_1 &=& 9 b \hspace{1.0cm} 
\bar{\gamma}_2 = 36 b^2 - \ell (\ell +1) + \lambda_1^2 + \lambda_1
\hspace{1.0cm} 
\bar{\gamma}_3 = 84 b^3 - 7 b \ell(\ell + 1)+\lambda_1^2(7b + 2 x_+) + 7b \lambda_1
                                                     \\   \nonumber
\bar{\gamma}_4 &=& 126 b^4 - \ell(\ell + 1)[21 b^2-x_+^2 - x_-^2]
                    + \lambda_1^2 [21 b^2 + 12 b x_+ + x_+^2 -2 x_-^2]
                    + \lambda_1[21 b^2 - x_+^2]
                                                     \\   \nonumber
\bar{\gamma}_5 &=& 126 b^5 - \ell(\ell + 1)[35 b^3 - 5 b(x_+^2+x_-^2)]
                    + \lambda_1^2 [35 b^3 + 30 b^2 x_+ + 5 b (x_+^2 - 2 x_-^2)
                       - 4 x_+ x_-^2]                 \\   \nonumber
& & + \lambda_1[35 b^3 - 5 b x_+^2 + 2 x_+ x_-^2]
                                                      \\   \nonumber
\bar{\gamma}_6 &=& 84 b^6 - \ell(\ell + 1)[35 b^4-10 b^2 (x_+^2+x_-^2)+ x_+^2 x_-^2]
                                                      \\   \nonumber
                 & & + \lambda_1^2 [35 b^4 + 40 b^3 x_+ + 10 b^2 (x_+^2 -2 x_-^2) 
                      -16 b x_+ x_-^2 - x_-^2 ( 2 x_+^2 - x_-^2)]
                                                      \\   \nonumber
                 & & + \lambda_1[35 b^4 - 10 b^2 x_+^2 + 8 b x_+ x_-^2 - x_-^4]
                                                      \\   \nonumber
\bar{\gamma}_7 &=& 36 b^7 - \ell(\ell+1)[21 b^5-10 b^3(x_+^2+x_-^2) +3b x_+^2 x_-^2]
                                                      \\   \nonumber
               & & + \lambda_1^2 [21 b^5 + 30 b^4 x_+ + 10 b^3(x_+^2 - 2 x_-^2)
                - 24 b^2 x_+ x_-^2 - 3 bx_-^2(2 x_+^2 - x_-^2) + 2 x_+ x_-^4]
                                                      \\   \nonumber  
              & & + \lambda_1[21 b^5 - 10 b^3 x_+^2 + 12 b^2 x_+ x_-^2 
                                 - 3 b x_-^4 - 2 x_+ x_-^2(x_+^2+x_-^2)]
                                                       \\   \nonumber
\bar{\gamma}_8 &=& 9 b^8 - \ell(\ell+1)[7 b^6-5 b^4(x_+^2+x_-^2)+3 b^2 x_+^2 x_-^2]
                                                       \\   \nonumber
            & & + \lambda_1^2 [7 b^6 + 12 b^5 x_+ + 5 b^4(x_+^2 -2 x_-^2)
                   - 16 b^3 x_+ x_-^2 - 3 b^2 x_-^2(2 x_+^2 - x_-^2)
                   + 4 b x_+ x_-^4 + x_+^2 x_-^4]       \\   \nonumber
           & & + \lambda_1[7 b^6 -5 b^4 x_+^2 + 8 b^3 x_+ x_-^2 - 3 b^2 x_-^4
                              -4 b x_+ x_-^2(x_+^2+x_-^2) + x_+^2 x_-^4]
                                                        \\   \nonumber
\bar{\gamma}_9 &=& b^9 - \ell(\ell+1)[b^7 - b^5(x_+^2+x_-^2)+ b^3 x_+^2 x_-^2]
                                                         \\   \nonumber
            & & + \lambda_1^2 [b^7 + 2 b^6 x_+ +b^5(x_+^2 -2 x_-^2)-4 b^4 x_+ x_-^2
                         -b^3 x_-^2(2 x_+^2 - x_-^2)+2b^2 x_+ x_-^4 + b x_+^2 x_-^4]
                                                         \\   \nonumber
           & & + \lambda_1[b^7 - b^5 x_+^2 + 2 b^4 x_+ x_-^2 - b^3 x_-^4 
                   -2 b^2 x_+ x_-^2(x_+^2+x_-^2)+ b x_+^2 x_-^4 + 2 x_+^3 x_-^4]
\end{eqnarray}
\end{appendix} 

\newpage
\begin{appendix}{\centerline{\bf Appendix B}}
\setcounter{equation}{0}
\renewcommand{\theequation}{B.\arabic{equation}}
The recursion relation in Eq. (\ref{blnhorizon1}) for $n=1$ is 
\begin{eqnarray}
\label{appena7}
& & d_{\ell,N} \left[(N + \lambda_1) (N + \lambda_1 - 1) \alpha_7 + 
                       (N + \lambda_1) \beta_7 + \gamma_9 \right]
                                                         \\   \nonumber
&+ & d_{\ell,N-1} \left[(N + \lambda_1 - 1) (N + \lambda_1 - 2) \alpha_6 
                     + (N + \lambda_1 - 1) \beta_6 + \gamma_8 \right]
                                                          \\   \nonumber
&+ & d_{\ell,N-2} \left[(N + \lambda_1 - 2) (N + \lambda_1 - 3) \alpha_5 
                     + (N + \lambda_1 - 2) \beta_5 + \gamma_7 \right]
                                                          \\   \nonumber
&+ & d_{\ell,N-3} \left[(N + \lambda_1 - 3) (N + \lambda_1 - 4) \alpha_4 
                     + (N + \lambda_1 - 3) \beta_4 + \gamma_6 \right]
                                                          \\   \nonumber
&+ & d_{\ell,N-4} \left[(N + \lambda_1 - 4) (N + \lambda_1 - 5) \alpha_3 
                     + (N + \lambda_1 - 4) \beta_3 + \gamma_5 \right]
                                                           \\   \nonumber
&+ & d_{\ell,N-5} \left[(N + \lambda_1 - 5) (N + \lambda_1 - 6) \alpha_2 
                     + (N + \lambda_1 - 5) \beta_2 + \gamma_4 \right]
                                                           \\   \nonumber
&+ & d_{\ell,N-6} \left[(N + \lambda_1 - 6) (N + \lambda_1 - 7) \alpha_1 
                     + (N + \lambda_1 - 6) \beta_1 + \gamma_3 \right]
                                                           \\   \nonumber
&+ & d_{\ell,N-7} \left[(N + \lambda_1 - 7) (N + \lambda_1 - 8) 
                          + 3 (N + \lambda_1 - 7) + \gamma_2 \right]
                                                           \\   \nonumber
&+ & d_{\ell,N-8} \gamma_1 + d_{\ell,N-9} = 0
\end{eqnarray}
where
\begin{eqnarray}
\label{appena8}
\alpha_1 &=& 9 x_+                    \hspace{0.6cm}  
 \alpha_2 = 34 x_+^2 - 2 x_-^2        \hspace{0.6cm}
 \alpha_3 = 70 x_+^3 - 14 x_+ x_-^2     \hspace{0.6cm}                 
 \alpha_4 = 85 x_+^4 -38 x_+^2 x_-^2 + x_-^4     
                                                            \\   \nonumber
\alpha_5 &=& 61 x_+^5 - 50 x_+^3 x_-^2 + 5 x_+ x_-^4     \hspace{0.5cm}    
 \alpha_6 = 24 x_+^6 -32 x_+^4 x_-^2 + 8 x_+^2 x_-^4       \hspace{0.5cm}
 \alpha_7 = 4 x_+^3 (x_+^2 - x_-^2)^2
                                                              \\   \nonumber
 \beta_1 &=& 26 x_+                        \hspace{.5cm}  
  \beta_2 = 80 x_+^2 + 4 x_-^2             \hspace{0.5cm}
  \beta_3 = 24 x_+ (6 x_+^2 - x_-^2)       \hspace{0.5cm}
  \beta_4 = 151 x_+^4 - 56 x_+^2 x_-^2 + x_-^4     
                                                               \\   \nonumber
 \beta_5 &=& 92 x_+^5 -64 x_+^3 x_-^2 + 4 x_+ x_-^4    \hspace{0.5cm}
  \beta_6 = 30 x_+^6 -36 x_+^4 x_-^2 + 6 x_+^2 x_-^4   \hspace{0.5cm} 
  \beta_7 = 4 x_+^3 (x_+^2 - x_-^2)^2    
                                                               \\   \nonumber
 \gamma_1 &=& 9 x_+                          \hspace{1.5cm}
 \gamma_2 = 36 x_+^2 - \ell(\ell+2)         \hspace{1.5cm}
 \gamma_3 = 84 x_+^3 - x_+ \ell(\ell+2)
                                                               \\   \nonumber
\gamma_4 &=& 126 x_+^4 + \ell(\ell+2)(4 x_+^2 + x_-^2)      \hspace{2.5cm}
 \gamma_5 = 126 x_+^5 - \ell(\ell+2) x_+ x_-^2          
                                                                \\   \nonumber
\gamma_6 &=& 84 x_+^6 - \ell(\ell+2)(13 x_+^4 + 3 x_+^2 x_-^2)  \hspace{1.8cm}
 \gamma_7 = 36 x_+^7 - \ell(\ell+2)(11 x_+^5 - 7 x_+^3 x_-^2)
                                                                \\   \nonumber
\gamma_8 &=& 9 x_+^8 - \ell(\ell+2)(2 x_+^6 -2 x_+^4 x_-^2)    \hspace{2.2cm}
 \gamma_9 = x_+^9
\end{eqnarray}

The recursion relation in Eq.(\ref{blasymp1}) for $n=1$ is  
\begin{eqnarray}
\label{appena9}
& &  \tau_{N (+)} [2i(N+1) + \tilde{\gamma}_1] 
     + \tau_{N-1 (+)} [N(N+1) - N \tilde{\beta}_1 + \tilde{\gamma}_2]
     + \tau_{N-2 (+)} [-(N-1) \tilde{\beta}_2 + \tilde{\gamma}_3]     
                                                                \\   \nonumber
& &  + \tau_{N-3 (+)} [(N-2)(N-1) \tilde{\alpha}_1 - (N-2) \tilde{\beta}_3 
                             + \tilde{\gamma}_4]
     + \tau_{N-4 (+)} [-(N-3) \tilde{\beta}_4 + \tilde{\gamma}_5]  
                                                                \\   \nonumber 
& &  + \tau_{N-5 (+)} [(N-4)(N-3) \tilde{\alpha}_2 -(N-4) \tilde{\beta}_5
                             + \tilde{\gamma}_6]
    + \tau_{N-6 (+)} [-(N-5) \tilde{\beta}_6 + \tilde{\gamma}_7]    
                                                                 \\   \nonumber
& & + \tau_{N-7 (+)} [(N-6)(N-5) \tilde{\alpha}_3 -(N-6) \tilde{\beta}_7 
                             + \tilde{\gamma}_8]
    + \tau_{N-8 (+)} [-(N-7) \tilde{\beta}_8 + \tilde{\gamma}_9]    
                                                                  \\   \nonumber
& & + \tau_{N-9 (+)} [(N-8)(N-7) \tilde{\alpha}_4 -(N-8) \tilde{\beta}_9
                             + \tilde{\gamma}_{10}] = 0
\end{eqnarray}
where $\tau_{0 (-)} = 1$, $\tau_{N (+)} = \tau_{N (-)}^*$ and
\begin{eqnarray}
\label{appena10}
\tilde{\alpha}_1 &=& -2(x_+^2 + x_-^2)                \hspace{0.5cm}
 \tilde{\alpha}_2 = x_+^4 + 4 x_+^2 x_-^2 + x_-^4      \hspace{0.5cm}  
 \tilde{\alpha}_3 = -2 x_+^2 x_-^2 (x_+^2 + x_-^2)      \hspace{0.5cm}
 \tilde{\alpha}_4 = x_+^4 x_-^4
                                                               \\  \nonumber
 \tilde{\beta}_1 &=& 2(1 + \lambda_1)            \hspace{0.5cm}
  \tilde{\beta}_2 = 2[x_+ \lambda_1 + 2i(x_+^2 + x_-^2)]     \hspace{0.5cm}
  \tilde{\beta}_3 = -2[(x_+^2 + x_-^2) + \lambda_1(x_+^2 + 2 x_-^2)]
                                                                \\  \nonumber
 \tilde{\beta}_4 &=& -2i[(x_+^4+4x_+^2 x_-^2+x_-^4)-i \lambda_1 x_+(x_+^2 + 2 x_-^2)]
                                        \hspace{2.0cm} 
  \tilde{\beta}_5 = 2 \lambda_1 x_-^2 (2 x_+^2 + x_-^2)
                                                                 \\  \nonumber
\tilde{\beta}_6 &=& 2 x_+ x_-^2[2i x_+(x_+^2 + x_-^2)+ \lambda_1 (2 x_+^2 + x_-^2)]
                                         \hspace{1.4cm}
\tilde{\beta}_7 = 2 x_+^2 x_-^2[(x_+^2 + x_-^2) - \lambda_1 x_-^2] 
                                                                 \\  \nonumber
\tilde{\beta}_8 &=& -2 x_+^3 x_-^4(i x_+ + \lambda_1)           \hspace{5.0cm} 
\tilde{\beta}_9 = -2 x_+^4 x_-^4
                                                                  \\  \nonumber
\tilde{\gamma}_1 &=& -2i(1 + \lambda_1)        \hspace{2.3cm}
\tilde{\gamma}_2 = -\frac{3}{4} -\ell(\ell + 2) + 2(x_+^2 + x_-^2)
                          + \lambda_1^2 + \lambda_1(1 - 2i x_+)   
                                                                   \\  \nonumber
\tilde{\gamma}_3 &=& 2i[(x_+^2 + x_-^2) - i\lambda_1^2 x_+ 
                                         + \lambda_1(x_+^2 + 2 x_-^2)]
                                                                  \\  \nonumber
\tilde{\gamma}_4 &=& \frac{1}{2} \bigg[-2(x_+^4+x_-^4) + 2 \ell(\ell+2)(x_+^2+x_-^2)
                     +(x_+^2+x_-^2) -8 x_+^2 x_-^2 + 2 \lambda_1^2 (x_+^2-2 x_-^2)
                                                                  \\  \nonumber
          & & \hspace{0.6cm}+ 2 \lambda_1[-x_+^2 + 2i x_+(x_+^2+2x_-^2)] \bigg]
              \hspace{0.8cm} 
\tilde{\gamma}_5 = -2 \lambda x_-^2[i(2 x_+^2 + x_-^2) - x_+(1 -2 \lambda_1)]
                                                                   \\  \nonumber
\tilde{\gamma}_6 &=& \frac{1}{4}  \bigg[(x_+^4+x_-^4) 
                              - 4 x_+^2 x_-^2[ \ell(\ell + 2)- 1 - 2(x_+^2+x_-^2)]
                               -4 x_-^2[ \lambda_1^2(2 x_+^2 - x_-^2)
                                                                    \\  \nonumber
& & \hspace{0.6cm} + \lambda_1[x_-^2 + 2ix_+(2 x_+^2 + x_-^2)] \bigg] \hspace{1.3cm}  
\tilde{\gamma}_7 = 2 x_+ x_-^2(\lambda_1 + i x_+) [\lambda_1 x_-^2 - (x_+^2+x_-^2)]
                          \hspace{0.6cm}
                                                                    \\  \nonumber
\tilde{\gamma}_8 &=& - \frac{1}{2} x_+^2 x_-^2 [3 x_+^2 + x_-^2[3 + 2 x_+^2 
                                        -2 \lambda_1^2 -2 \lambda_1(1+2i x_+)]]
                                                                  \\  \nonumber
\tilde{\gamma}_9 &=& 2 x_+^3 x_-^4 (\lambda_1 + i x_+)       \hspace{3.9cm} 
\tilde{\gamma}_{10} = \frac{5}{4} x_+^4 x_-^4
\end{eqnarray}

The recursion relation in Eq.(\ref{power}) for $n=1$ is
\begin{eqnarray}
\label{appena11}
& & D_N [N(N-1) \bar{\alpha}_9] 
  + D_{N-1}[(N-1)(N-2) \bar{\alpha}_8+(N-1)\bar{\beta}_9]
                                                                 \\  \nonumber
&+& D_{N-2} [(N-2)(N-3) \bar{\alpha}_7 + (N-2) \bar{\beta}_8 + \bar{\gamma}_9]
+ D_{N-3} [(N-3)(N-4) \bar{\alpha}_6 + (N-3) \bar{\beta}_7 + \bar{\gamma}_8]
                                                 \\   \nonumber
&+& D_{N-4} [(N-4)(N-5) \bar{\alpha}_5 + (N-4) \bar{\beta}_6 + \bar{\gamma}_7]
+ D_{N-5} [(N-5)(N-6) \bar{\alpha}_4 + (N-5) \bar{\beta}_5 + \bar{\gamma}_6]
                                                 \\   \nonumber
&+& D_{N-6} [(N-6)(N-7) \bar{\alpha}_3 + (N-6) \bar{\beta}_4 + \bar{\gamma}_5]
+ D_{N-7} [(N-7)(N-8) \bar{\alpha}_2 + (N-7) \bar{\beta}_3 + \bar{\gamma}_4]
                                                 \\   \nonumber
&+& D_{N-8} [(N-8)(N-9) \bar{\alpha}_1 + (N-8) \bar{\beta}_2 + \bar{\gamma}_3]
+ D_{N-9} [(N-9)(N-10)  + (N-9) \bar{\beta}_1 + \bar{\gamma}_2]
                                                 \\   \nonumber
&+& D_{N-10} \bar{\gamma}_1 + D_{N-11} = 0
\end{eqnarray}
where
\begin{eqnarray}
\label{appena12}
\bar{\alpha}_1 &=& 9b                 \hspace{1.0cm} 
\bar{\alpha}_2 = 36 b^2 -2(x_+^2 + x_-^2)
                                      \hspace{1.0cm} 
\bar{\alpha}_3 = 14 b[6b^2 -(x_+^2+x_-^2)]
                                                \\   \nonumber
\bar{\alpha}_4 &=& 126 b^4 + (x_+^4+x_-^4) + 4 x_+^2 x_-^2 -42 b^2(x_+^2+x_-^2)                                                \\   \nonumber
\bar{\alpha}_5 &=& b[126 b^4 - 70 b^2 (x_+^2 + x_-^2)
                          + 5(x_+^4 + x_-^4 + 4x_+^2 x_-^2)]
                                                 \\   \nonumber
\bar{\alpha}_6 &=& 2 [42 b^6 - 35 b^4(x_+^2+x_-^2)-x_+^2 x_-^2(x_+^2+x_-^2)
                   + 5 b^2(x_+^4 + x_-^4 + 4 x_+^2 x_-^2)]
                                                 \\   \nonumber
\bar{\alpha}_7 &=& 2 b[18 b^6 - 21 b^4(x_+^2+x_-^2)-3x_+^2 x_-^2(x_+^2+x_-^2)
                     +5 b^2(x_+^4 + x_-^4 + 4 x_+^2 x_-^2)]
                                                 \\   \nonumber
\bar{\alpha}_8 &=& [b^4 - b^2(x_+^2+x_-^2)+x_+^2 x_-^2] [9 b^4 +x_+^2 x_-^2
                                -5 b^2(x_+^2+x_-^2)]
                                                 \\   \nonumber
\bar{\alpha}_9 &=& b[b^4 - b^2(x_+^2+x_-^2)+x_+^2 x_-^2]^2
                                                 \\   \nonumber
\bar{\beta}_1 &=& 3 + 2 \lambda_1       \hspace{1.0cm}
\bar{\beta}_2 = 2[12b + \lambda_1(8b + x_+)]
                                                 \\   \nonumber
\bar{\beta}_3 &=& 4(21b^2 - x_+^2 - x_-^2) 
                  + 2 \lambda_1(28b^2 +7b x_+ - x_+^2 - 2 x_-^2)
                                                  \\   \nonumber 
\bar{\beta}_4 &=& 2 \bigg[ 12[7b^3 - b(x_+^2+x_-^2)] 
                    + \lambda_1[56b^3+21b^2 x_+ -(6b+x_+)(x_+^2+2x_-^2)] \bigg] 
                                                  \\   \nonumber
\bar{\beta}_5 &=& 210b^4 -60b^2(x_+^2+x_-^2)+ x_+^4+x_-^4 + 4x_+^2x_-^2
                                                  \\   \nonumber
             & & + 2 \lambda_1[70b^4+35b^3 x_+-5b(3b+x_+)(x_+^2+2 x_-^2)
                                              + x_-^2(2x_+^2+x_-^2)]
                                                   \\   \nonumber
\bar{\beta}_6 &=& 4[42b^5-20b^3(x_+^2+x_-^2)+b(x_+^4+x_-^4+4x_+^2x_-^2)]
                                                   \\   \nonumber
              & & + 2 \lambda_1[56b^5+35b^4x_+ - 10b^2(2b+x_+)(x_+^2+2x_-^2) 
                                               + x_-^2(4b+x_+)(2x_+^2+x_-^2)]
                                                    \\   \nonumber
\bar{\beta}_7 &=& 6[14b^6-10b^4(x_+^2+x_-^2)+b^2(x_+^4+x_-^4+4x_+^2x_-^2)]
                                                   \\   \nonumber
              & & + 2 \lambda_1[28b^6+21b^5x_+ - 5b^3(3b+2x_+)(x_+^2+2x_-^2)
                                     +3bx_-^2(2b+x_+)(2x_+^2+x_-^2)-x_+^2x_-^4]
                                                    \\   \nonumber
\bar{\beta}_8 &=& 4[6b^7-6b^5(x_+^2+x_-^2)+b^3(x_+^4+x_-^4+4x_+^2x_-^2)]
                                                    \\   \nonumber 
              & & +2 \lambda_1[8b^7 + 7b^6 x_+-b^4(6b+5x_+)(x_+^2+2x_-^2)
                         +b^2x_-^2(4b+3x_+)(2x_+^2+x_-^2)-x_+^2x_-^4(2b+x_+)] 
                                                     \\   \nonumber
\bar{\beta}_9 &=& (b^2-x_+^2)(b^2-x_-^2) \bigg[3b^4-b^2(x_+^2+x_-^2)-x_+^2x_-^2
                            + \lambda_1[2b^4+2b^3x_+-2b^2x_-^2-2bx_+x_-^2] \bigg]
                                                      \\   \nonumber
\bar{\gamma}_1 &=& 9 b         \hspace{1.0cm} 
\bar{\gamma}_2 = 36 b^2 - \ell (\ell + 2) + \lambda_1^2 + 2 \lambda_1
                                                      \\   \nonumber
\bar{\gamma}_3 &=& 84 b^3 - 7b \ell (\ell + 2) + \lambda_1^2 (7b+2x_+)
                          + \lambda_1(14b+x_+)
                                                       \\   \nonumber
\bar{\gamma}_4 &=& 126 b^4 - \ell(\ell + 2)[21 b^2-x_+^2 - x_-^2]
                                                       \\   \nonumber
               & &  + \lambda_1^2 [21 b^2 + 12 b x_+ + x_+^2 -2 x_-^2]
                    + 2 \lambda_1[21 b^2+3bx_+-x_+^2 - x_-^2]
                                                     \\   \nonumber
\bar{\gamma}_5 &=& 126 b^5 - \ell(\ell + 2)[35 b^3 - 5 b(x_+^2+x_-^2)]
                    + \lambda_1^2 [35 b^3 + 30 b^2 x_+ + 5 b (x_+^2 - 2 x_-^2)
                       - 4 x_+ x_-^2]                 \\   \nonumber
& & + \lambda_1[70 b^3 + 15 b^2 x_+ - 10b(x_+^2+x_-^2)- x_+^3]
                                                      \\   \nonumber
\bar{\gamma}_6 &=& 84 b^6 - \ell(\ell + 2)[35 b^4-10 b^2 (x_+^2+x_-^2)+ x_+^2 x_-^2]
                                                      \\   \nonumber
                 & & + \lambda_1^2 [35 b^4 + 40 b^3 x_+ + 10 b^2 (x_+^2 -2 x_-^2) 
                      -16 b x_+ x_-^2 - x_-^2 ( 2 x_+^2 - x_-^2)]
                                                      \\   \nonumber
      & & + 2 \lambda_1[35 b^4 + 10 b^3 x_+ -10b^2(x_+^2+x_-^2) -2bx_+^3 + x_+^2x_-^2]
                                                      \\   \nonumber
\bar{\gamma}_7 &=& 36 b^7 - \ell(\ell+2)[21 b^5-10 b^3(x_+^2+x_-^2) +3b x_+^2 x_-^2]
                                                      \\   \nonumber
               & & + \lambda_1^2 [21 b^5 + 30 b^4 x_+ + 10 b^3(x_+^2 - 2 x_-^2)
                - 24 b^2 x_+ x_-^2 - 3 bx_-^2(2 x_+^2 - x_-^2) + 2 x_+ x_-^4]
                                                      \\   \nonumber  
              & & + \lambda_1[42 b^5 +15b^4 x_+ - 20 b^3 (x_+^2+x_-^2) - 6 b^2 x_+^3
                                                + 6b x_+^2x_-^2 - x_+ x_-^4]
                                                       \\   \nonumber
\bar{\gamma}_8 &=& 9 b^8 - \ell(\ell+2)[7 b^6-5 b^4(x_+^2+x_-^2)+3 b^2 x_+^2 x_-^2]
                                                       \\   \nonumber
            & & + \lambda_1^2 [7 b^6 + 12 b^5 x_+ + 5 b^4(x_+^2 -2 x_-^2)
                   - 16 b^3 x_+ x_-^2 - 3 b^2 x_-^2(2 x_+^2 - x_-^2)
                   + 4 b x_+ x_-^4 + x_+^2 x_-^4]       \\   \nonumber
           & & + \lambda_1[14 b^6 +6b^5 x_+ - 10 b^4 (x_+^2+x_-^2) -4 b^3 x_+^3
                             + 6b^2 x_+^2 x_-^2 -2b x_+ x_-^4]
                                                          \\   \nonumber
\bar{\gamma}_9 &=& b^9 - \ell(\ell+2)[b^7 - b^5(x_+^2+x_-^2)+ b^3 x_+^2 x_-^2]
                                                         \\   \nonumber
            & & + \lambda_1^2 [b^7 + 2 b^6 x_+ +b^5(x_+^2 -2 x_-^2)-4 b^4 x_+ x_-^2
                         -b^3 x_-^2(2 x_+^2 - x_-^2)+2b^2 x_+ x_-^4 + b x_+^2 x_-^4]
                                                         \\   \nonumber
           & & + \lambda_1[2b^7 +b^6 x_+ -2 b^5(x_+^2 +x_-^2) - b^4 x_+^3 
                                  +2b^3 x_+^2 x_-^2 -b^2 x_+ x_-^4 + x_+^3 x_-^4]
\end{eqnarray}

\end{appendix}

%\begin{figure}
%\caption[fig1]{Plot of the partial absorption cross sections 
%for massless scalar case. The black lines represent the partial 
%absorption cross sections computed
%by analytic solutions and their analytic continuations. The red lines 
%are result of Ref.[16]. The fact $\sigma_0 = 4 \pi$ at $k = 0$ indicates the
%universality of the low-energy absorption cross section for S-wave.
%The peak points of $\sigma_{\ell}$ are given in Table I.}
%\end{figure}
%\vspace{0.4cm}
%\begin{figure}
%\caption[fig2]{The $r_{\ast}$-dependence of the effective potential
%(\ref{effective1}) for massless
%case. The potential makes a barrier, which separates the asymptotic and
%near-horizon regions. The barrier heights are given in Table I.}
%\end{figure}

\end{document}